%% file: main.tex
\documentclass[aps,nofootinbib,prd,twocolumn,superscriptaddress,10pt]{revtex4-2}

\usepackage[dvipsnames]{xcolor}
\definecolor{HBlue}{RGB}{0,0,122}
\definecolor{HRed}{RGB}{173,42,26}
\usepackage[colorlinks=true,linkcolor=black, urlcolor=HBlue,citecolor=HRed]{hyperref}
\usepackage{amsmath}
\DeclareMathOperator*{\argmin}{arg\,min}
\usepackage{amssymb}
\usepackage{booktabs}
\usepackage[capitalise]{cleveref}
\usepackage{enumitem}
\usepackage{fontawesome5}
\crefname{section}{Sec.}{Secs.}
\usepackage{graphicx}
\usepackage{siunitx}
\usepackage{comment}
\DeclareSIUnit\bar{bar}
\DeclareSIUnit\hr{h}
\DeclareSIUnit\ueV{\micro\eV}
\usepackage{tabularx}
\usepackage{xspace}

\input{macros}

\input{journal_macros}
\newcommand{\updated}[1]{#1}

\begin{document}

\title[\hyperlsw]{Ultimate light-shining-through-a-wall experiments to establish QCD axions as the dominant form of dark matter}

\author{Sebastian Hoof\orcid{0000-0002-3517-2561}}
\email{hoof@pd.infn.it}
\affiliation{Dipartimento di Fisica e Astronomia ``Galileo Galilei'', Universit\`a degli Studi di Padova, Via F.\ Marzolo~8, 35131 Padova, Italy}
\affiliation{Istituto Nazionale di Fisica Nucleare -- Sezione di Padova,
Via F.\ Marzolo~8, 35131 Padova, Italy}
\author{Joerg Jaeckel\orcid{0000-0002-6038-4785}}
\email{jjaeckel@thphys.uni-heidelberg.de}
\affiliation{Institut f\"ur Theoretische Physik, Universit\"at Heidelberg, Philosophenweg 16, 69120 Heidelberg, Germany}
\author{Giuseppe Lucente\orcid{0000-0003-1530-4851}}
\email{lucenteg@slac.stanford.edu}
\affiliation{Kirchhoff-Institut f\"ur Physik, Universit\"at Heidelberg, Im Neuenheimer Feld 227, 69120 Heidelberg, Germany}
\affiliation{Institut f\"ur Theoretische Physik, Universit\"at Heidelberg, Philosophenweg 16, 69120 Heidelberg, Germany}

\date{\today}

\begin{abstract}
\updated{Establishing the axion as the dark matter (DM) particle after a haloscope discovery typically requires follow-up experiments to break the degeneracy between the axion's coupling to photons and its local DM abundance.}
Given that a discovery would justify more significant investments, \updated{we explore} the prospects of ambitious light-shining-through-a-wall (LSW) setups to \updated{probe} the QCD~axion band.
Leveraging the excellent mass determination in haloscopes, we show how to design \updated{LSW experiments} with \updated{lengths on the order of \SI{100}{\km} and} suitably aligned magnetic fields \updated{with apertures of around \SI{1}{\m}} to reach well-motivated axion models \updated{across up to four orders of magnitude in mass}.
\updated{Beyond} presenting a concrete plan for post-discovery experimental efforts, we briefly discuss complementary experiments and future directions beyond LSW experiments.
\end{abstract}

\maketitle

\section{Introduction\label{sec:intro}}

QCD~axions~\cite{Wilczek:1977pj,Weinberg:1977ma} and axion-like particles~(ALPs)~\cite{Kim:1986ax,1002.0329} are not only a solution to the strong~$CP$ problem~\cite{Peccei:1977hh,Peccei:1977ur}, but also excellent dark matter~(DM) candidates~\cite{Preskill:1982cy,Abbott:1982af,Dine:1982ah,Turner:1983he,Turner:1985si,1201.5902}.
The extensive campaign of ongoing experimental searches (see, e.g., Ref.~\cite{1801.08127} for an overview) targets a broad range of axion models~\cite[e.g.][]{2003.01100} and, importantly, realistic DM QCD~axion models.

An axion discovery would be a monumental achievement and mark the first milestone in an experimental campaign focused on determining axion properties, establishing them as the dominant form of DM, and illuminating their connection to the strong~$CP$ problem.
Incidentally, a discovery in a haloscope~\cite{Sikivie:1983ip} leaves a parameter degeneracy between the axion-photon coupling \gagg and their fraction \afrac of the local DM density \rholoc.\footnote{When encountering an axion minicluster, it may be possible to break this degeneracy with a haloscope experiment alone~\cite{2307.11871}.}
In fact, a single experiment is rarely capable of self-consistently determining multiple axion parameters.
Possible exceptions are helioscope searches~\cite{Sikivie:1983ip} for axions with masses $\ma \sim \SI{10}{\meV}$~\cite{1811.09278,1811.09290}.
However, helioscope data alone will likely yield sizable parameter uncertainties, limiting our ability to infer details of the underlying model, such as evidence for its connection to QCD.

Given these challenges, a mass determination would significantly aid the development of customized follow-up experiments and justify funding for more ambitious designs.
We thus explore the potential for such follow-up experiments, starting from the information available from a haloscope discovery, which facilitate the construction of a light-shining-through-a-wall (LSW) experiment~\cite{Anselm:1985obz,VanBibber:1987rq} that is guaranteed to provide a complementary probe of the newly discovered particle.\footnote{It is presently difficult search for non-DM QCD axion models via couplings other than \gagg in laboratory-based searches, with the ARIADNE experiment~\cite{1403.1290} as a potential exception in case of an extra source of $CP$ violation~\cite[e.g.][]{2010.03889}.}

The rationale behind our approach is twofold.
First, the haloscope measures the product $\gagg^2\afrac\rholoc$.
By using the maximum allowed value,\footnote{The probability of accidentally passing through a local overdensity, such as an axion minicluster, can be reduced by measuring for a sufficiently long time.} i.e.\ $\afrac = 1$, we can determine a minimal value for \gagg, which sets a target to guarantee a successful LSW measurement.
Second, an accurate \ma measurement allows us to arrange a specific magnetic field configuration, providing optimal sensitivity at that mass.
This is important because standard LSW experiments cannot achieve the sensitivity required to reach the QCD axion band with currently available technology.
Modified LSW setups can overcome this limitation, although typically only for a limited mass range, underscoring the importance of knowing \ma.

Based on this approach, we identify the QCD~axion parameter space where our proposed ``\hyperlsw'' experiments can test axions.
The resulting independent measurement of \gagg, in combination with the haloscope measurement, determines the local axion fraction \afrac.
Consequently, we can conclude whether axions are the dominant component of the local DM density or merely a smaller fraction.

Although the idea of decisively testing the QCD axion parameter space has been discussed, and initial proposals for coordinated efforts have been made~\cite[e.g.][]{Graham:2015ouw,2203.14923},\footnote{Indeed, Ref.~\cite{Graham:2015ouw} contains a suggestion for an ambitious LSW experiment (``JURA'') based on FCC magnets.} we want to push LSW experiments to the boundaries of current feasibility and to explore how important questions beyond an initial discovery can be addressed.

Our strategy is outlined in \cref{sec:strategy}, while we summarize LSW setups and how to optimize them in \cref{sec:lsw_physics} and \cref{sec:optimizing_the_setup}, respectively.
The combined \hyperlsw sensitivity is discussed in \cref{sec:results}, where we also consider possible extensions and other use cases.
We conclude with a summary in \cref{sec:conclusions} and provide additional information and supplementary computations as appendices.

\section{General strategy\label{sec:strategy}}

The cornerstone of our search strategy is a discovery in a haloscope experiment~\cite{Sikivie:1983ip}, which is particularly promising due to its excellent capabilities to determine \ma and the axion's direct connection to DM.
To break the parameter degeneracy between \gagg and \afrac, we will rely on an LSW experiment~\cite{Anselm:1985obz,VanBibber:1987rq}, which can self-consistently determine \gagg without depending on the axion's DM nature.
We identify the alternating-magnet LSW designs from Refs.~\cite{VanBibber:1987rq,1009.4875} as a promising\updated{, flexible} option to enhance the sensitivity \updated{at specific mass values across a wide range of axion masses.}

\subsection{Haloscope discovery\label{sec:haloscope}}

Haloscope experiments can detect axions from the local DM population around Earth.
We chiefly consider \emph{resonant} searches, which are tunable and allow a superb determination of \ma, in addition to inferring $\gagg^2\afrac\rholoc$.
The feasibility of these searches has been demonstrated by the RBF~\cite{DePanfilis:1987dk} and UF~\cite{Hagmann:1990tj} Collaborations, and the long-running ADMX experiment~\cite{astro-ph/0310042,0910.5914,1804.05750,1910.08638,2110.06096}.
More recently, the CAPP Collaboration~\cite{2001.05102,2008.10141,2012.10764,2206.08845,2207.13597,2210.10961,2308.09077} and others~\cite{1706.00209,1803.03690,1903.06547,2008.01853,2012.09498,2104.13798,2110.14406,2211.02902,2203.12152,2205.05574,2208.12670,2301.09721} have also presented competitive limits.

Haloscopes have also been recognized as tools to further study local DM properties -- including the DM velocity distribution or DM substructure~\cite{1701.03118,1711.10489,1806.05927,2307.11871}.
Indeed, the sensitivity to $\gagg^2\afrac\rholoc$ depends on the shape of the local velocity distribution at the time of the discovery.
However, once a signal is found, its time series can be analyzed to infer the local DM structure, as discussed in the works cited above.
Substructure, such as streams or voids, can affect the line shape of the axion signal at high resolution, but the underlying Maxwellian shape and axion mass can still be inferred~\cite{2311.17367}.

The key point for our purposes is the high precision in determining \ma after a haloscope discovery.
For instance, for $\ma \sim \si{\ueV}$ and a Maxwellian halo, the expected uncertainty of \ma is well below $\Delta \ma \lesssim \SI{1}{\Hz} = \SI{4e-15}{\eV}$ after a year of observations~\cite[Fig.~2]{1701.03118}.
For intermediate \ma, this level of precision is more than sufficient for our purposes, as we explicitly demonstrate in \cref{sec:optimizing_the_setup} and \cref{sec:Fexpansion}.
However, as we will see, the upper end of masses detectable with the more advanced versions of \hyperlsw would require a better mass resolution, making this case more challenging.

\subsection{Light-shining-through-a-wall follow-up}\label{sec:lsw}

Our follow-up experiment of choice is an LSW experiment~\cite{Anselm:1985obz,VanBibber:1987rq}, which both generates and detects axions via~\gagg.
Such setups have been realized by a number of collaborations such as ALPS~\cite{0905.4159,1004.1313}, OSQAR~\cite{0712.3362,1110.0774,1506.08082}, and others~\cite{10.1007/BF01474722,Cameron:1993mr,hep-ph/0605250,0707.1296,0710.3783,0806.2631,1310.8098,1510.08052} (see Ref.~\cite{1011.3741} for a review), and the currently ongoing ALPS~II experiment~\cite{1302.5647} will significantly extend the LSW reach into previously untested parameter space.

\updated{Light-shining-through-a-wall experiments are typically not sensitive to QCD axion models since they would have to be extremely long when using currently available magnetic field strengths.}
Even then, they stop gaining in sensitivity at a length scale $\sim 2\pi\omega/\ma^2$, where $\omega$ is the angular photon frequency.
This is because the axion and the photon wave increasingly go out of phase, leading to destructive interference \updated{and posing a major issue for QCD axion masses (see \cref{sec:lsw_physics})}.

\updated{Various modifications of the basic setup} have been proposed \updated{to improve the sensitivity to larger masses and thereby in particular QCD~axion models}. 
These include optical resonators~\cite{Hoogeveen:1990vq,0907.5387}, phase shift plates~\cite{0706.0693}, alternating magnet configurations~\cite{VanBibber:1987rq,1009.4875}, superconducting radio frequency cavities~\cite{1904.07245}, magnetic field profiles~\cite{1910.09973,2108.01486}, or axion magnetic resonance~\cite{2308.10925}.
\updated{In this work, we will utilize optical resonators, since they give an overall sensitivity improvement.
In addition we employ specific magnet configurations that allow to resonantly enhance the sensitivity across a wide range of axion masses.
The other proposals are typically more limited in their applicability, although conducting a more careful analysis might prove beneficial (see \cref{sec:conclusions}).}

\updated{Note that the downside of resonantly enhancing the signal is a rather time-consuming, difficult re-arrangement of the setup when scanning across a range of axion masses.
Our assumption of a haloscope discovery, and hence a precise measurement of the axion mass, eliminates this issue.}

\section{Summary of LSW physics}\label{sec:lsw_physics}

Let us summarize the basic formulae governing the axion-photon inter-conversion in LSW experiments -- particularly for setups with multiple magnets, gaps, and possibly changing field orientations.
Our discussion is largely based on Ref.~\cite{1009.4875}.

Consider axions and photons propagating along the $z$~axis inside a magnetic field of magnitude $B$ with spatial variation $|f(z)| \leq 1$ along a perpendicular axis.
For \nm magnets of total length $L \equiv \nm \ell$, the square of the axion-photon interconversion probability is given by
\begin{equation}
    \pag^2 = \frac{\omega^2}{\omega^2 - \ma^2} \left(\frac{\gagg B L}{2}\right)^4 |F|^4 ,
    \label{eq:p_agamma}
\end{equation}
with form factor
\begin{align}
    F \equiv \frac{1}{L} \int_{0}^{L} \! \dd z \; f(z) \, \ee^{\ii q z} \label{eq:integralF}
\end{align}
and momentum transfer
\begin{align}
    q \equiv \nref \,\omega - \sqrt{\omega^2 - \ma^2} \simeq (\nref - 1) \, \omega + \frac{\ma^2}{2\omega} , \label{eq:momentum_transfer}
\end{align}
where \nref is the refractive index inside the magnetic fields.
The approximation in \cref{eq:momentum_transfer} is valid for $\ma \ll \omega$ and matches the well-known expression for vacuum ($\nref = 1$)\updated{, which we will assume throughout this work (see \cref{sec:gas_filling} for issues related to gas-filled setups)}.
\begin{widetext}
Allowing for gaps of size $\Delta$ between the individual magnets of length $\ell$, grouped in \ngr groups \updated{of alternating polarity with} \nset magnets each, Ref.~\cite{1009.4875} found that the form factor can be written as
\begin{equation}
    F_{\nset,\ngr}(x;\delta) = \frac{\sinc(x)}{\ngr \nset} \, \frac{\tan\left(\nset \, y \right)}{\sin(y)}
    \left\{ \begin{array}{ll}
        \sin\left(\nset\ngr \, y\right) & \text{if \ngr is even} \\
        \cos\left(\nset\ngr \, y\right) & \text{if \ngr is odd}
    \end{array}\right\} , \label{eq:generalF}
\end{equation}
where we defined $\sinc(x) \equiv \sin(x)/x$, $x \equiv q\ell/2$, $y \equiv x (1+\delta)$, and $\delta \equiv \Delta/\ell$.
\end{widetext}
The ``gapless'', single-magnet limit of \cref{eq:generalF}, i.e.\ $\delta = 0$ and $\nm = \ngr \, \nset = 1$, reproduces the expected limits:
\begin{align}
    F_{\nset,\ngr}(x;\delta) &\overset{\delta = 0}{=} \; \frac{\tan\left(\nset \, x\right)}{\nset\ngr \, x}
    \left\{ \begin{array}{ll}
        \sin\left(\nset\ngr \, x\right) & \text{if \ngr is even} \\
        \cos\left(\nset\ngr \, x\right) & \text{if \ngr is odd}
    \end{array} \right\} \nonumber \\
    \; &\overset{\nm = 1}{=} \sinc(x) \equiv F_{1,1}(x) . \label{eq:limitingF}
\end{align}

The expected number of signal photons $\mathcal{S}$ is given by
\begin{equation}
    \mathcal{S} \equiv \mtx{\varepsilon}{eff} \, \frac{\plaser \, \tau}{\omega} \, \mtx{\beta}{g} \, \mtx{\beta}{r} \, \pag^2 , \label{eq:n_photons}
\end{equation}
where the effective efficiency $\mtx{\varepsilon}{eff}$ includes the mode overlap factor between the photon and axion modes~(cf.\ Ref.~\cite{1009.4875}) and the detector efficiency, \plaser is the laser power, and $\tau$ is the measurement time.
The quantities $\mtx{\beta}{g}$ and $\mtx{\beta}{g}$ are respectively the boost factors of the generation and regeneration parts of the LSW experiment, i.e.\ the boost factors before and after the wall.
In each case, the boost factor $\beta$ is given by
\begin{equation}
    \beta^{-1} \equiv \beta_0^{-1} + \ee^{-\zeta} , \label{eq:boost}
\end{equation}
where $\ee^{-\zeta}$ defines the clipping losses and $\beta_0^{-1}$ summarize all other loss sources (see \cref{sec:optics} for details).

The clipping losses in \cref{eq:boost} turn out to be a leading factor in limiting the reach for LSW experiments, as discussed in \cref{sec:optimizing_the_setup}.
Magnets with sufficiently large aperture could, in principle, prevent these losses and allow for the construction of LSW experiments that are hundreds of kilometers long -- albeit by introducing the challenges and costs associated with large apertures.
However, at such length scales another challenge becomes relevant: since LSW experiments need to follow a straight line, the curvature of Earth becomes a relevant, limiting factor.
To illustrate this point, consider digging a straight tunnel of length $2L$, which emerges from Earth's surface on both ends.
The depth $d$ in the middle of the tunnel is
\begin{equation}
    d = R_\oplus - \sqrt{R_\oplus^2 - L^2} \simeq \frac{L^2}{2R_\oplus} = \SI{785}{\m} \left(\frac{L}{\SI{100}{\km}}\right)^2 , \label{eq:earth_curvature}
\end{equation}
where $R_\oplus \approx \SI{6370}{\km}$ is Earth's approximate radius.
Note that $d$ in \cref{eq:earth_curvature} also corresponds to the maximum height of the support structures needed if \hyperlsw were to be constructed on or, more accurately, tangential to Earth's surface.
Therefore, and to maintain sufficient mechanical stability, a tunnel is preferable over surface construction.
The deepest mine in the world, the Mponeng mine in South Africa, operates at a depth of about \SI{4}{\km}~\cite{mponeng_mine}, corresponding to an experiment of total length $2L \approx \SI{450}{\km}$.
These operations have to deal with high temperatures and other logistical challenges, but can nonetheless inform a limit on the possible length.

\section{Optimizing \hyperlsw setups for QCD axion models}\label{sec:optimizing_the_setup}

\begin{figure}
    \centering
    \includegraphics[width=3.375in]{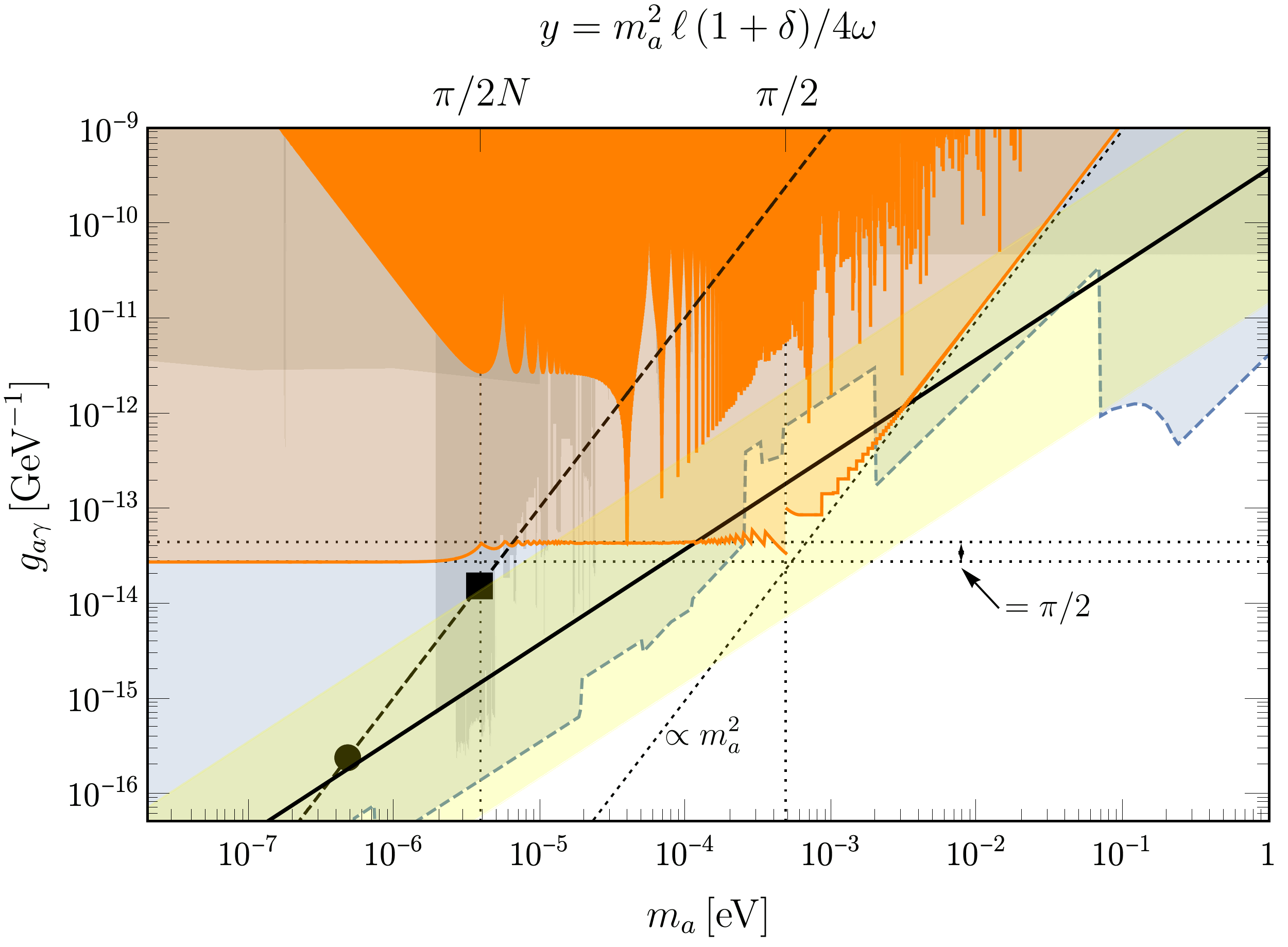}
    \caption{Properties and scaling relations of LSW experiments. The combined addressable parameter space (light orange) for the S1~setup from \cref{tab:bench_par} is compared to a single realization for $\ma = \SI{40}{\ueV}$ (dark orange). We also show the maximal sensitivity for an LSW setup ($B = \SI{10}{\tesla}$; dashed black line) of total length of \SI{200}{\km} (black square) and $2R_\oplus$ (black dot), the QCD axion band (yellow region), KSVZ model (solid black line), various constraints (gray region), and projected haloscope sensitivity (light blue region, dashed line). See \cref{fig:exp} for references.}
    \label{fig:explanatory}
\end{figure}

The main challenge for a long LSW experiment is that axion and photon waves start interfering destructively beyond a length $\sim 2\pi \omega/\ma^2$, corresponding to $x \sim \pi/2$ for a single magnet.
Increasing the length beyond this point does not increase the sensitivity because the form factor decreases, thus limiting the experiment's sensitivity.
This is illustrated in \cref{fig:explanatory}, where the dashed black line shows the $3\sigma$ median discovery sensitivity \cite[Eq.~(97)]{1007.1727} that can be reached with a single magnet of $B = \SI{10}{\tesla}$ and $L \sim 2\pi \omega/\ma^2$, assuming the S1~setup in \cref{tab:bench_par} for all other parameters.
The black square in \cref{fig:explanatory} corresponds to a total length of $2L = \SI{200}{\km}$, while the black dot is for $2L = 2R_\oplus$.

We can compare the LSW sensitivity to QCD~axion models, which have an axion-photon coupling \gagg of
\begin{equation}
    \gagg = \frac{\alphaEM}{2\pi\fax} \, \cagg = \frac{\alphaEM}{2\pi\fax} \left|E/N - \caggzero\right| , \label{eq:gagg}
\end{equation}
with anomaly ratio $E/N$, axion decay constant \fax, model-independent term $\caggzero = \num{1.92(4)}$~\cite{1511.02867}, and fine-structure constant $\alphaEM \approx 1/137$.
The solid black line in \cref{fig:explanatory} shows the Kim--Shifman--Vainshtein--Zakharov (KSVZ) model~\cite{Kim:1979if,Shifman:1979if} ($E/N = 0$), while the yellow QCD~axion band spans values $\cagg \in [\num{0.0722}, \, \num{17.3}]$.
These correspond to the central 95\% region from a Monte Carlo simulation of equally-weighted, joint catalogs~\cite{2107.12378,2302.04667} of theoretically preferred~\cite{1610.07593,1705.05370} Dine--Fischler--Srednicki--Zhitnitsky (DFSZ)~\cite{Zhitnitsky:1980tq,Dine:1981rt} and KSVZ axion models (available on Zenodo~\cite{dfsz_catalogue}), where we included the uncertainty of \caggzero via a normal distribution, ${\caggzero \sim \mathcal{N}(\mu,\sigma^2)=\mathcal{N}(1.92,0.04^2)}$. 

From \cref{fig:explanatory}, it is evident that no realistic, standard LSW experiment with $B \lesssim \SI{10}{\tesla}$ can probe the KSVZ benchmark model.
\updated{The alternating-magnet design, however, can resonantly enhance LSW sensitivity, potentially reaching the QCD~axion model band.
In what follows, we detail how to achieve the optimal sensitivity for a given value of \ma in terms of the number of magnets $N$, grouping (given by \nset), and gap size $\Delta$.

For instance, the opaque orange region in \cref{fig:explanatory} represents the optimal sensitivity for $\ma = \SI{40}{\ueV}$, while the transparent orange region is the combined experimental reach of all \hyperlsw experiments that could be constructed for a given \ma in that mass range.
To obtain the total experimental reach, we have to distinguish three different cases, the ``low'', ``intermediate'', and ``high'' mass region.

We discuss these cases separately in \cref{sec:opt:m0,sec:opt:m1,sec:opt:m2}, but the main outcome can broadly be summarized as follows:
\begin{itemize}
    \item The form factor in \cref{eq:generalF} is independent of \ma in the ``low-mass'' regime, $y \lesssim \pi/2 N$, and so is the LSW experiment's sensitivity.
    The magnets polarizations are aligned and gaps are minimal.
    \item Coherent conversion in a single magnet becomes impossible for ``large'' masses, $y \gtrsim \pi/2$.
    The optimal configuration requires fully-alternating magnets and minimal gaps, leading to a sensitivity scaling of $\gagg \sim \ma^2$ (see \cref{sec:opt:m2} for a derivation of the sensitivity scaling and \cref{sec:opt:m2plus} for further optimization in the transition region).
    \item In the ``intermediate-mass'' regime, $\pi/2 N \lesssim y \lesssim \pi/2$, the optimal setups in some sense interpolate between the previous two cases, with the number of magnets per group approximately following \cref{eq:master}.
    The best sensitivity in this region is approximately constant, and it is typically a factor $\pi/2 \approx 1.6$ lower than at ``low'' masses.
\end{itemize}}

\subsection{The low-mass region}\label{sec:opt:m0}

For $\ma \to 0$, \cref{eq:generalF} goes to $0$ for even \ngr, and to $1/\ngr$ for odd \ngr.
The optimal setup is thus $\ngr = 1$ and does not depend on the relative gap size~\rgap.
Since the conversion probability in \cref{eq:p_agamma} is proportional to $L^2$, it is optimal to chain many magnets of length $\ell$ with their $B$~fields aligned in the same direction.
We refer to this as the ``fully-aligned'' setup.

The argument above ignores the clipping losses in \cref{eq:boost}, which effectively limit the total length of the LSW setup.
The balance between longer setups and clipping losses results in an optimal length \zopt, which generally needs to be computed numerically~\cite{1009.4875} (see Appendix~\ref{sec:optics} for more details).
\updated{An approximate value for \zopt for the production (or regeneration) part of the experiment can however be obtained following Ref.~\cite[Eq.~(33)]{1009.4875}.
For instance, for $\beta_0 = \num{e5}$, one finds that}
\begin{equation}
    \zopt \approx \SI{94.2}{\km} \left(\frac{\SI{1064}{\nm}}{\lambda}\right) \left(\frac{a}{\SI{1.3}{\m}}\right)^2 , \label{eq:zopt}
\end{equation}
where $\lambda$ is the laser wavelength and $a$ the aperture diameter of the magnet.
Since \zopt refers to the length of only one part of the experiment, the total length of a symmetric setup would be $2 \times \zopt$.

To determine the optimal number of magnets per part of the LSW setup, we choose
\begin{equation}
    \nm = \argmin_{\nm \in \mathbb{N}} \left| z_\nm - \zopt \right| = \mathrm{round}\left(\frac{\zopt + \mtx{\Delta}{min}}{\ell + \mtx{\Delta}{min}} \right) , \label{eq:nmopt}
\end{equation}
where $z_\nm = \nm \ell + (\nm - 1) \Delta$ and $\mtx{\Delta}{min}$ is the smallest allowed gap size.
For the setup~S1 from \cref{tab:bench_par}, \cref{eq:zopt} gives $\zopt \approx \SI{94}{\km}$, while \cref{eq:nmopt} results in $\nm \approx \num{15700}$.

\subsection{The intermediate-mass region}\label{sec:opt:m1}

Once $\nm y \sim \pi/2$, the last factor in \cref{eq:generalF} starts oscillating, and we need to consider configurations beyond the fully-aligned setup.
This condition corresponds to $\ma \sim \SI{4}{\ueV}$ for setup~S1, as shown in \cref{fig:explanatory}.

According to Ref.~\cite{1009.4875}, the maxima of \cref{eq:generalF} are close to the poles of the tangent, located at
\begin{equation}
    x_k = \frac{\left(1 + 2k\right) \pi}{2 \nset (1 + \delta)} \quad \text{for $k \in \mathbb{N}_0$} . \label{eq:tangent_poles}
\end{equation}
The largest maximum occurs for $k = 0$, where we find that (see \cref{sec:Fmaxima} for more details)
\begin{equation}
    \left|\frac{F_{\nset,\ngr}(x;\delta)}{F_{1,1}(x;\delta)}\right| \to \left|\frac{1}{\nset \sin\left(\pi/2\nset\right)} \right| \simeq \frac{2}{\pi} \quad (x \to x_0) , \label{eq:maxF}
\end{equation}
assuming that $\nset \gg 1$ for the last approximation.
\Cref{eq:maxF} indicates that, unless $\nset = 1$, the sensitivity is at most a factor of $2/\pi$ lower than the optimal sensitivity in the massless limit (see \cref{sec:opt:m0}).
This ratio is also indicated in \cref{fig:explanatory}.

In particular, we can achieve an optimized setup by matching $\nset \, y = \pi/2$.
This condition also connects well with the low-mass and high-mass regions since ${1 \leq \nset \leq \nm}$, and it can be compactly written as
\begin{equation}
    \nset = \min\left\{\nm, \max\left\{1, \mathrm{round}\left(\pi/2\,y\right)\right\}\right\} ,\label{eq:master}
\end{equation}
leading to optimal configurations for $1/N \leq 2y/\pi \leq 1$, as shown by the vertical dotted lines in \cref{fig:explanatory}.
More details on the optimization strategy for the intermediate-mass region are provided in \cref{sec:generalstrategy}.

For setups other than the fully-aligned one, we have to consider the width of the mass region at which we achieve optimal sensitivity.
This becomes evident from \cref{fig:explanatory}, where the maximal sensitivity of a single configuration (solid orange region) is achieved in a rather narrow mass region.
As shown in \cref{sec:Fexpansion}, its width is
\begin{equation}
\label{eq:width}
    \frac{\Delta \ma}{\ma} \simeq \frac{\sqrt{6}}{4 x_k (1+\delta)} \frac{1}{\nm} \approx \frac{\sqrt{6}}{2(1+2k)\pi}\frac{\nset}{\nm}.
\end{equation}
As $y_{k} = x_{k}(1+\delta) \lesssim \pi/2$ we find that, for the intermediate-mass region, we require a relative mass resolution better than
\begin{equation}
    \frac{\Delta \ma}{\ma} \sim \frac{0.4}{\nm} \gtrsim \num{e-5} .
\end{equation}
This is larger than the physical width of the axion peak in haloscope experiments and therefore problematic.

\subsection{The high-mass region}\label{sec:opt:m2}

For $y > \pi/2$, the loss of coherence is unavoidable -- even for a single magnet -- since the condition in \cref{eq:tangent_poles} cannot be met anymore when $k = 0$.
For our S1~setup, the critical mass is $\ma \sim \SI{0.5}{\meV}$, which is a factor of $\sqrt{N} \sim \mathcal{O}(100)$ higher than the ${\ma \sim \SI{4}{\ueV}}$ threshold for the intermediate-mass region.

According to \cref{eq:master}, when $y > \pi/2$, $\nset = 1$ and thus $\ngr = \nm$.
In this ``fully-alternating'' setup of magnets, the poles of the tangent in \cref{eq:tangent_poles} now correspond to maxima of the form factor in \cref{eq:generalF}, as already noted in Ref.~\cite{1009.4875}.
The form factor in this setup reduces to that of a single magnet of length $\ell$ when $x \to x_k$.
\Cref{eq:p_agamma,eq:n_photons} then imply that the sensitivity can be close to \nm times that of a single magnet of length $\ell$ as long as $x \approx x_k$.
We can achieve this by adjusting \rgap to match $x = x_k$ in \cref{eq:tangent_poles} with the smallest possible~$k$.

\begin{figure}
    \centering
    \includegraphics[width=3.375in]{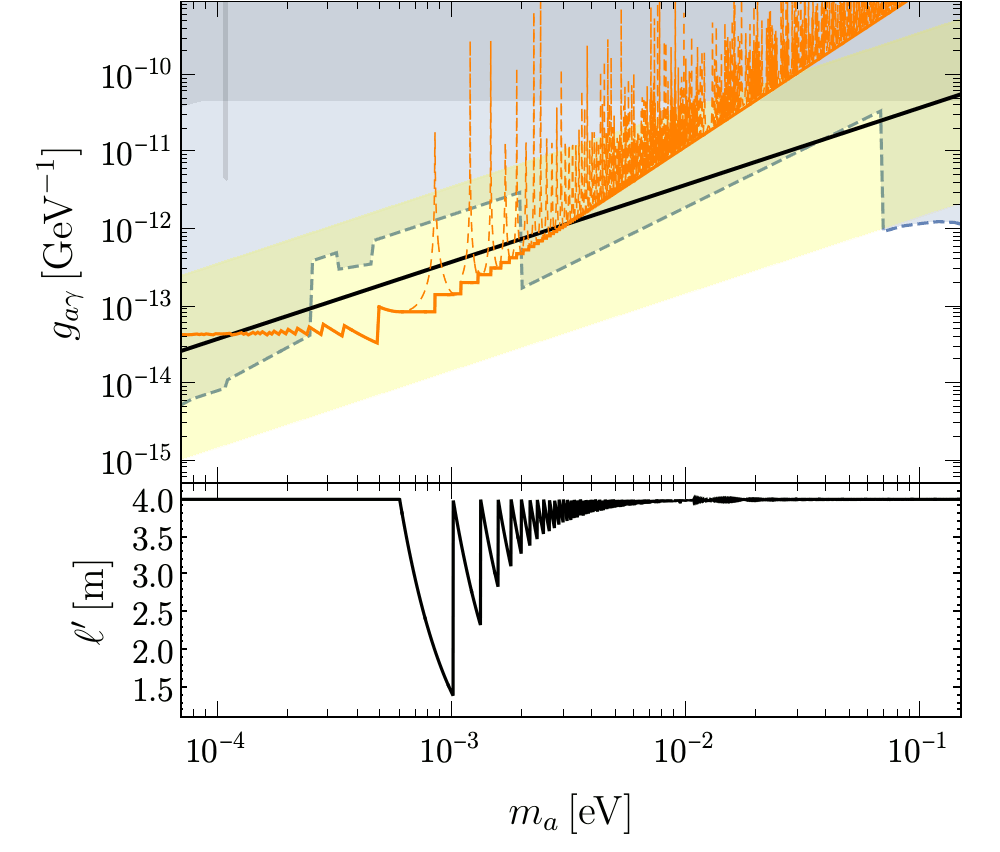}
    \caption{Further optimization of the high-mass region. The baseline (dashed orange line) and improved sensitivity (solid orange line; top panel) for the S1 setup are shown after shortening the magnet lengths (bottom panel).
    We also show the QCD axion band (yellow region), KSVZ model (solid black line), various constraints (gray region), and projected haloscope sensitivity (light blue region, dashed line). See \cref{fig:exp} for references.}
    \label{fig:M1_improve}
\end{figure}

However, this approach leaves gaps in the sensitivity related to the zeros of the form factor of a single magnet of length~$\ell$.
This is demonstrated via the dashed orange line in the upper panel of \cref{fig:M1_improve}, which shows the sensitivity in the high-mass region the S1~setup, linked to the single-magnet form factor $F_{1,1}(x) = \sinc(x)$.
Specifically, $\mathcal{S} \propto \gagg^4\,\sinc(x)^4$, and the sensitivity worsens as $\gagg \propto 1/\sinc(x) \sim \ma^2$.
This explains the sensitivity gaps, related to the zeros of $F_{1,1}(x)$ at $x = k\pi$ for $k \in \mathbb{N}$.
We discuss how to close these gaps in \cref{sec:opt:m2plus}.

Again, we have to consider the required mass resolution for the haloscope measurement.
\Cref{eq:width} now implies a more stringent requirement because $k$, and $x_{k}$, are now significantly larger,
\begin{equation}
    \frac{\Delta\ma }{\ma} \sim  \num{e-6} \left(\frac{\num{1e4}}{\nm}\right) \left(\frac{\SI{6}{\m}}{\ell(1+\delta)}\right) \left(\frac{\omega}{\SI{1}{\eV}}\right) \left(\frac{\SI{2.5}{\meV}}{\ma}\right)^2 .
\end{equation}
As a result, our ``S-type'' setup should not encounter serious problems up to masses of about \SI{2.5}{\meV}.
At higher masses, however, a better resolution is required.
As long as a mass resolution of $\Delta\ma/\ma \sim \numrange{e-9}{e-8}$ can be achieved, cf.\ \cite[Fig.~2]{1701.03118}, we may probe masses in the region $\ma \sim \SIrange{25}{75}{\meV}$, which is comparable to the highest masses that can be reached with our setup (see \cref{sec:benchmark_parameters}).
We note, however, that the mass resolution quoted above was estimated for an axion mass of \SI{1}{\ueV}~\cite{1701.03118}, and reaching the same resolution at higher masses may be more challenging.
To alleviate this issue, one may consider different detection techniques discussed in Ref.~\cite{breadtalk}, or potentially obtain an even more precise mass measurement with a dedicated follow-up haloscope setup after a discovery.

\subsection{Further optimization for high masses}\label{sec:opt:m2plus}

As discussed in \cref{sec:opt:m2}, the zeros in $F_{1,1}(x)$ cannot be avoided by changing~\rgap.
One approach to address this issue, which has been successfully employed in the past~\cite[e.g.][]{1004.1313}, involves using a buffer gas to alter the refractive index, thereby changing the momentum transfer in \cref{eq:momentum_transfer}.
However, due to the increased absorption and scattering losses over the long lengths of \hyperlsw setups, this solution may be problematic (see \cref{sec:gas_filling}).

Instead, we can exploit the relation $x \propto \ell$ and adjust the length of the magnets.
From \cref{eq:maxF}, we know that the optimal form factor for the fully-alternating configuration reduces to $F_{1,1}(x)$ at $x_k$.
We therefore may shorten $\ell \mapsto \ell^\prime < \ell$, such that $\ell^\prime$ maximizes $F_{1,1}(x)$.
To be at the maximum of the fully-alternating configuration, the condition from \cref{eq:tangent_poles} must be satisfied, implying that
\begin{equation}
    \ell (1 + \delta) = \ell^\prime (1 + \delta^\prime) .
\end{equation} 
The maxima of the form factor $|\sinc(x^\prime)|$ are close to
\begin{equation}
    x^\prime_{k^\prime} = \frac{\pi}{2} + k^\prime \pi \quad \text{for $k^\prime \in \mathbb{N}$} , \label{eq:max_sinc}
\end{equation}
where $x^\prime \equiv q\,\ell^\prime / 2$.
For each \ma, we find the largest possible value of $\ell^\prime \leq \ell$ satisfying \cref{eq:max_sinc}, corresponding to the largest maximum of the form factor close to $x_k$.

While shortening of the magnet to the $k^\prime = 1$ maximum would be optimal, doing so may be technically challenging, and we thus stay in the vicinity of the $k$th maximum to avoid large changes to~$\ell$.
More details and caveats on the optimization strategy for high masses can be found in \cref{sec:generalstrategy}.

The length adjustment described above and shown in the lower panel of \cref{fig:M1_improve} leads to the the sensitivity shown in the upper panel as a solid orange line.
The shortest length required to avoid the first zero of $F_{1,1}$ is $\ell^\prime \approx \SI{1.5}{\m}$ -- i.e.\ almost a factor~3 smaller than $\ell$ -- at $\ma \approx \SI{1}{\meV}$, while $\ell^\prime = \ell$ for larger \ma.

\section{Results and discussion}\label{sec:results}

Before constructing \hyperlsw, it is crucial to ensure that we can measure \afrac with sufficient precision to meet our science goals.
Recall that the haloscope signal scales as $\gagg^2 \afrac \rholoc$, where the average local DM density $\rholoc = \SIrange{0.3}{0.6}{\GeV/\cm^3}$~\cite{1404.1938,2012.11477} has sizeable systematic uncertainties.

A haloscope signal might in principle result from a local overdensity, such as axion miniclusters, meaning that $\afrac > 1$.
However, we can exclude this possibility by observing the signal spectrum for a longer period, beyond these usually brief and rare encounters.\footnote{An overdensity from a ``bound object'' would likely also be visible as a very narrow feature in the temporal spectrum~\cite[e.g.][]{1701.03118}.}
The ``worst case'' scenario for \hyperlsw is therefore when \gagg takes the smallest possible value for the allowed signal, corresponding to $\afrac = 1$ and \rholoc = \SI{0.6}{\GeV/\cm^3} (possibly multiplied with a safety factor to account for more exotic halo models).
That minimal value of \gagg is the target threshold, which does not depend on the axion model.

The rationale above provides a model-independent ``no-lose theorem'', i.e., a clear answer to the question of whether or not it makes sense to build \hyperlsw.
Also note that if the target \gagg is larger than the technical sensitivity limit presented in \cref{fig:exp}, we can apply the cost-saving measures from \cref{sec:timeline_and_costs}.
Assuming a QCD~axion, we can also estimate $\afrac \leq 1$ for a given cosmological scenario, which we will briefly discuss in \cref{sec:other_uses}.

Let us now estimate the sensitivity of a \hyperlsw setup.
Since we assume the axion to already have been discovered, it is not meaningful to define the sensitivity in terms of a median expected limit or discovery reach.
Instead, we follow the suggestion of Ref.~\cite[Sec.~40]{10.1093/ptep/ptac097} and define the figure-of-merit 
\begin{equation}
    \Phi \equiv \frac{\mathcal{S}}{\sqrt{\mathcal{S} + \mathcal{B}}} , \label{eq:fom}
\end{equation}
whose inverse $\Phi^{-1}$ measures the expected relative uncertainty of the signal~$\mathcal{S}$ in \cref{eq:n_photons} for a background~$\mathcal{B}$.
Since $\mathcal{S} \propto \gagg^4$, the precision for measuring \gagg is then approximately
\begin{equation}
    \Pi \equiv \frac{\Delta \gagg}{\gagg} = \frac{1}{4\Phi} .
\end{equation}
This allows us to find the critical signal threshold $\mtx{\mathcal{S}}{crit}$ to achieve a desired precision $\Pi$, 
\begin{equation}
    \mtx{\mathcal{S}}{crit} = \frac{1}{32\,\Pi^2} \left(1 + \sqrt{1 + 64\,\mathcal{B} \, \Pi^2}\right) .
\end{equation}

The choice of $\Pi$ depends on our science goals.
We can assume that the haloscope provides a precise estimate for $\gagg^2 \afrac \rholoc$ due to the high statistical significance required for a detection, and \hyperlsw can also potentially estimate $\gagg^4$ very precisely.
The uncertainty on \afrac will thus mainly arise from the systematic spread in \rholoc estimates, which roughly lie within a factor of~1.3 within the intermediate value of $\rholoc = \SI{0.45}{\GeV/\cm^3}$.
Potential future reductions in the systematic uncertainties of \rholoc justify aiming for a precision of around 20\% for \afrac ($\Pi = 10\%$), aligning the statistical uncertainty of \afrac with the systematic uncertainty of \rholoc.

While going beyond that level is not indicated as far as \afrac is concerned, an alternative science target could be to measure \gagg at the level of theoretical uncertainty for the QCD axion.
This may allow us to infer the corresponding $E/N$, narrowing down the underlying axion model from the available model catalogs~\cite{2107.12378,2302.04667}.
The limiting factor is the theoretical uncertainty from axion-meson mixing, encoded in the uncertainty on \caggzero in \cref{eq:gagg}.
For instance, for the KSVZ model ($E/N = 0$), we would require a target precision of $\Pi = 2\%$.

Given the effort required to build \hyperlsw, we use $\Pi = 2\%$ as a more ambitious target.

\subsection{Sensitivity for benchmark setups}\label{sec:benchmark_parameters}

\begin{table*}
\caption{Benchmark setups for \hyperlsw experiments\updated{, inspired by the future MADMAX and ongoing ALPS~II experiments (see main text for details)}. The magnet designs (indicated by numbers~``1'' and~``2'') are characterized by magnetic field strength $B$, aperture diameter $a$, magnet length~$\ell$, and minimal gap between magnets $\mtx{\Delta}{min}$. The optics and detector depend on the laser power \plaser and wavelength $\lambda$, the intrinsic cavity boost factors in the generation ($\mtx{\beta}{g}$) and regeneration ($\mtx{\beta}{r}$) parts, overall efficiency~$\mtx{\varepsilon}{eff}$, dark count rate~$b$, and measurement time~$\tau$\updated{, for which we adopt more conservative, standard (``S'') and more optimistic (``O'') values}. We also include the optimal/maximal \hyperlsw length ($2\,\zopt$) and the required signal strength $\mtx{\mathcal{S}}{crit}$ to reach our science goals.\label{tab:bench_par}}
\begin{tabular}{lS[table-format=2]S[table-format=1.1]S[table-format=2.1]S[table-format=1.1]S[table-format=3]S[table-format=0e1]S[table-format=0e1]S[table-format=4]S[table-format=1.1]S[table-format=4]S[table-format=0e-1]lS[table-format=3.1] }
    \toprule
    Setup & \multicolumn{1}{c}{$B$ [\si{\tesla}]} & \multicolumn{1}{c}{$a$ [\si{\m}]} & \multicolumn{1}{c}{$\ell$ [\si{\m}]} & \multicolumn{1}{c}{$\mtx{\Delta}{min}$ [\si{\m}]}  & \multicolumn{1}{c}{\plaser [\si{\W}]} & \multicolumn{1}{c}{$\mtx{\beta}{g}$} & \multicolumn{1}{c}{$\mtx{\beta}{r}$} & \multicolumn{1}{c}{$\lambda$ [\si{\nm}]} & \multicolumn{1}{c}{$\mtx{\varepsilon}{eff}$} & \multicolumn{1}{c}{$\tau$ {[\si{\hr}]}} & \multicolumn{1}{c}{$b$ \si{[\s^{-1}]}}& \multicolumn{1}{c}{$2 \,\zopt$ [\si{\km}]} & \multicolumn{1}{c}{$\mtx{\mathcal{S}}{crit}$} \\
    \midrule
    S1 & 9 & 1.3 & 4.0 & 2.0  & 3 & 1e5 & 1e5 & 1064 & 0.9 & 100 & 1e-4 & {$2\times94$} & 186.4 \\ 
    S2 & 11 & 1.8 & 10.0 & 3.0 & 3 & 1e5 & 1e5 & 1064 & 0.9 & 100 & 1e-4 & {$2\times181$} & 186.4 \\
    \midrule
    O1 & 9 & 1.3 & 4.0 & 2.0  & 300 & 1e5 & 1e6 & 1064 & 0.9 & 5000 & 1e-6 & {$2\times79$} & 172.5 \\ 
    O2 & 11 & 1.8 & 10.0 & 3.0 & 300 & 1e5 & 1e6 & 1064 & 0.9 & 5000 & 1e-6 & {$2\times152$} & 172.5 \\
    \bottomrule
\end{tabular}
\end{table*}

In addition to the S1~setup, already introduced in previous sections, we provide further benchmark setups in \cref{tab:bench_par}.
In particular, the letters ``S'' and ``O'' respectively denote configurations with more conservative (``standard'') and more optimistic experimental parameters, while the numbers~1 and~2 indicate different magnet designs.
The values for ``type~1''  and ``type~2'' magnets are inspired by prototype magnets for the MADMAX experiment~\cite{10.1109/TASC.2020.2989478,10.1109/TASC.2023.3315201}.
We also impose a minimal gap size that is somewhat larger than $a$ to ensure a good magnetic field quality within the gaps (see, e.g., Ref.~\cite{10.1007/978-3-319-18317-6}).
Setups with ``type~2'' magnets assume both a larger aperture diameter~$a$ and length~$\ell$.
They thus lead to longer experimental setups (cf.\ \zopt column in \cref{tab:bench_par}) and better sensitivity in the low-mass and intermediate-mass regions compared to ``type-1'' magnets.
However, due to their larger~$\ell$, ``type-2'' magnets do not improve the sensitivity in the high-mass regime since
 the condition $y > \pi/2$ is already fulfilled at smaller masses.

The non-magnet parameters that we consider are largely inspired by the achieved or projected capabilities of the ALPS~II \updated{experiment~\cite[e.g.\ Table~1]{1302.5647}}.

\begin{figure*}
    \centering
    \includegraphics[width=0.75\textwidth]{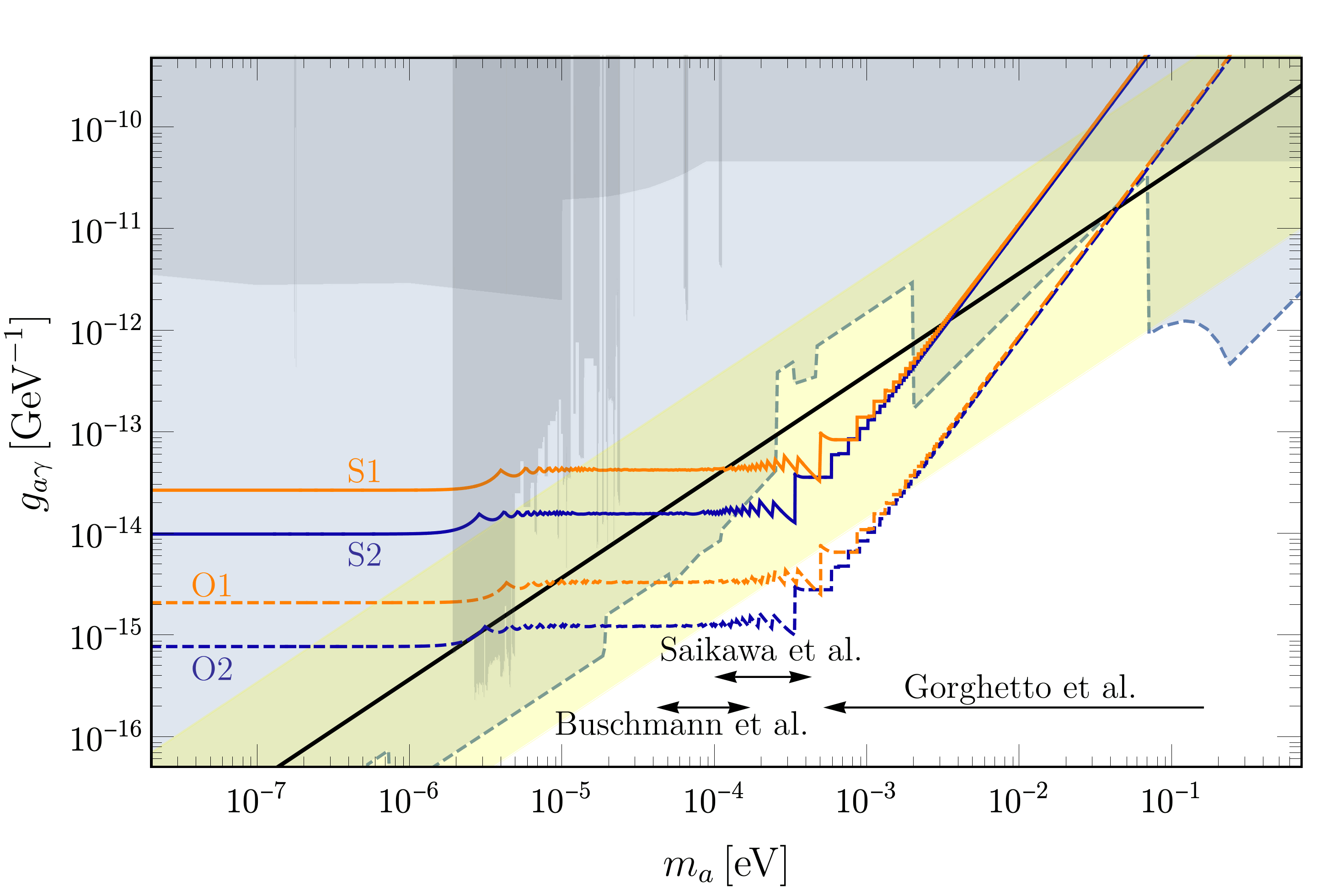}
    \caption{Reach of \hyperlsw in terms of measuring \gagg to a precision of 2\% or better. Solid and dashed lines indicate the ``S''  and ``O'' setups, while orange and blue lines correspond to ``type~1'' and ``type~2'' magnets (see \cref{tab:bench_par}). The QCD axion band (yellow region; defined in \cref{sec:optimizing_the_setup}), KSVZ model (solid black line), various constraints (gray region)~\cite{DePanfilis:1987dk,Wuensch:1989sa,Hagmann:1990tj,Hagmann:1996qd,0910.5914,1406.6053,1706.00209,1803.03690,1804.05750,1901.00920,1903.06547,1910.08638,1911.05772,2001.05102,2008.01853,2008.10141,2012.09498,2012.10764,2104.13798,2110.06096,2110.10262,2110.14406,2203.12152,2205.05574,2206.08845,2206.12271,2207.03102,2207.13597,2208.12670,2209.09917,2210.10961,2210.10961,2211.02902,2301.09721,2304.07505,2309.00351,2310.00904,2312.11003,2402.12892,2403.07790,2405.19393} and projected haloscope sensitivity (light blue region, dashed line)~\cite{1612.08296,1706.00209,1803.11455,1904.11872,2003.10894,2111.12103,2204.13781,2206.02980,2303.03997,2306.05934,2306.17243} are also shown (adapted from Ref.~\cite{axionlimits}). The black arrows indicate various predicted QCD axion mass ranges from cosmological simulations~\cite{1806.04677,1906.00967,2108.05368,2007.04990,2401.17253} up to the mass limit from hot dark matter overproduction~\cite{2205.07849}.}
\label{fig:exp}
\end{figure*}

In \cref{fig:exp}, we show the sensitivity for all benchmark setups.
They can achieve QCD axion sensitivity (with respect to our target precision) over a significant mass range and, in particular, KSVZ axions sensitivity in the following mass ranges:
\begin{align}
\label{eq:massrange}
    \ma^\text{KSVZ} &\in [0.11,\,3.4]\,\text{\si{\meV} for setup S1} , \nonumber \\
    \ma^\text{KSVZ} &\in [0.002,\,45]\,\text{\si{\meV} for setup O2} . \nonumber
\end{align}

As in previous figures, we indicate the region that can be probed by existing (gray regions) and proposed (light blue region delimited by a dashed line) haloscope experiments.
The overlapping regions of the projected sensitivity for future haloscopes and our benchmark setups are the currently most intriguing ones for envisioning \hyperlsw due to a possible future a discovery, and hence the required precise mass measurement.
\updated{The KSVZ mass ranges above can further be compared to future bounds from cosmology, which could limit $\ma \lesssim \SI{40}{\meV}$~\cite{2310.08169}.}
Regarding these mass values, we highlight the existance of the ``meV gap'' in haloscope searches for large parts of the QCD~axion band.
Closing this gap is technologically challenging but would be required to realize the full potential for \hyperlsw.

\subsection{Potential issues and extensions}\label{sec:potential_issues}

Having demonstrated the general feasibility of \hyperlsw under idealized conditions, we now discuss some of the potential issues for a realistic implementation.

\subsubsection{Experimental uncertainties}

Reaching the QCD~axion band relies on resonantly enhancing the LSW setup by specific magnet arrangements, in particular at larger \ma.
We thus have to ensure the following:
\begin{itemize}
    \item The width of the resonance peaks is larger than the precision of the haloscope measurement.
    \item The unavoidable errors from magnet imperfections and positioning do not spoil the resonance.
\end{itemize}

In \cref{sec:Fexpansion}, we estimate the width of the resonance peaks and find that, thanks to the superb mass resolution of haloscopes, sub-meV masses are generally unproblematic.
For larger masses, we require $\Delta \ma/\ma \sim \numrange{e-9}{e-8}$, as discussed in \cref{sec:opt:m2}.

In \cref{sec:errors}, we compute analytical estimates and perform Monte Carlo simulations to estimate the systematic shift and uncertainty on the form factor $|F|$ due to various sources of uncertainties.
We find that random errors in the absolute magnet positioning of order $\mathcal{O}(\SI{1}{\cm})$ become problematic for $\ma \gtrsim \SI{4}{\meV}$.
At larger masses, we would need to reduce the positioning errors.
For instance, to reach $\ma = \SI{40}{\meV}$, \cref{eq:F_with_errors} requires positioning errors of order \SI{100}{\micro\m}.

Regarding misalignment and size differences in the magnetic field profiles, issues arise for $\ma \gtrsim \SI{0.5}{\meV}$.
Apart from also reducing the uncertainties, we can measure and thus characterize the magnetic field profiles of all magnets and subsequently compensate any undesired phase shifts to achieve the target sensitivity.
While this increases the setup's complexity, we deem such efforts to be realistically achievable.

\subsubsection{Cost and construction time estimates}\label{sec:timeline_and_costs}

Achieving the highest possible sensitivity in \hyperlsw experiments incurs significant construction costs, primarily driven by tunneling and magnet construction.

The cost of a single magnet is roughly proportional to the stored magnetic energy in the magnet volume, which scales as $B^2 \ell a^2$.
Since the setups in \cref{tab:bench_par} are inspired by the magnet designs for the future MADMAX experiment, we can use the related design studies to estimate the costs.
Given the anticipated unit cost of about 0.01\,GEUR (1\,GEUR = one billion euro)~\cite{10.1109/TASC.2020.2989478,10.1109/TASC.2023.3315201}, the total cost for \nm magnets scales as 
\begin{equation}
    \mtx{C}{magn} \sim 10\,\text{GEUR} \left(\frac{B}{\SI{9}{\tesla}}\right)^2 \left(\frac{\ell\,a^2}{\SI{6.8}{\m^3}}\right)\left(\frac{\nm}{1000}\right) . \label{eq:cost_mag}
\end{equation}

Tunneling cost can, in principle, be compared by normalizing to the excavated volume.
However, it does not necessarily scale linearly with the cross sectional area~\cite[chart~G.1]{2010_infrastructure_report} or length~\cite[chart~G.2]{2010_infrastructure_report}.
The former is relevant for \hyperlsw due to the large aperture of the magnets, while the latter is due to planning costs and other expenses, which disproportionally affect shorter tunnels of length $z \lesssim \SI{25}{\km}$~\cite[chart~7.5]{2020_bts_hyperloop}.
Since we anticipate much longer setups, we may assume a linear scaling with the total length of the experiment.
As a reference, the cost for a future ``Hyperloop'' long-distance transport tunnel with a \SI{6.5}{\m} diameter has been estimated to cost around 8.4\,GEUR/\SI{100}{\km}~\cite{2020_bts_hyperloop,10.1680/jcien.20.00022}.\footnote{Unless stated otherwise, all costs are adjusted to January 2024 price levels.}
This is in line with the cost of the Gotthard Base Tunnel (7.6\,GEUR/\SI{100}{\km}; total length of all tunnels: \SI{152}{\km}; total cost of 9.5\,GCHF at 1998 price levels)~\cite{2019_NEAT_Kommission} or estimated cost of the Brenner Base Tunnel (4.6\,GEUR/\SI{100}{\km}; total length of all tunnels: \SI{230}{\km}; total cost of 10.5\,GEUR at 2023 levels).\footnote{\label{note:bbt}See the project website at \url{https://www.bbt-se.com/en/tunnel/project-overview/} for information and updates on the Brenner Base Tunnel.}
Thus, for an experiment of total length $Z$, we estimate the tunneling cost to be
\begin{equation}
    \mtx{C}{tunnel}\sim 10\,\text{GEUR} \left(\frac{Z}{\SI{100}{km}}\right) . \label{eq:cost_tunnel}
\end{equation}

Ignoring the costs of the optical components and operations, the two benchmark setups would cost 
\begin{equation}
    S_1: \; \num{200}\,\text{GEUR}, \quad S_2: \;  \num{1000}\,\text{GEUR}. \nonumber
\end{equation}

These are, of course, rather crude estimates, and more refined estimates will depend on the experiment's location and other factors.

\subsubsection{\updated{Potential cost reductions}}\label{sec:cost_reductions}

\updated{The tentatively high costs of \hyperlsw may be justified by additional uses for the facility, which we briefly discuss in \cref{sec:other_uses}.
However, there is also potential for cost savings due to mass production of the magnets, which we did not take into account in our estimates.}

Moreover, if the haloscope measurement suggests a relatively high \gagg target, we could scale down \hyperlsw while still achieving its primary physics goals.
Specifically, one could reduce the number of magnets or use a smaller aperture $a$ to save money since, according to \cref{eq:zopt}, the sensitivity scales as ${\gagg \propto (\nm\ell)^{-1} \sim a^{-2}}$ if $N\ell \sim \zopt$.
If the magnet cost dominates, $\mtx{C}{magn} > \mtx{C}{tunnel}$, and the total cost approximately scales as $B^2a^4$, implying $\gagg \propto \mtx{C}{magn}^{-1/2}$.
For $\mtx{C}{tunnel} > \mtx{C}{magn}$, costs will scale as $\mtx{C}{tunnel}\propto z \sim a^2$, implying $\gagg \propto \mtx{C}{tunnel}^{-1}$.

\begin{figure}
    \centering
    \includegraphics[width=3.375in]{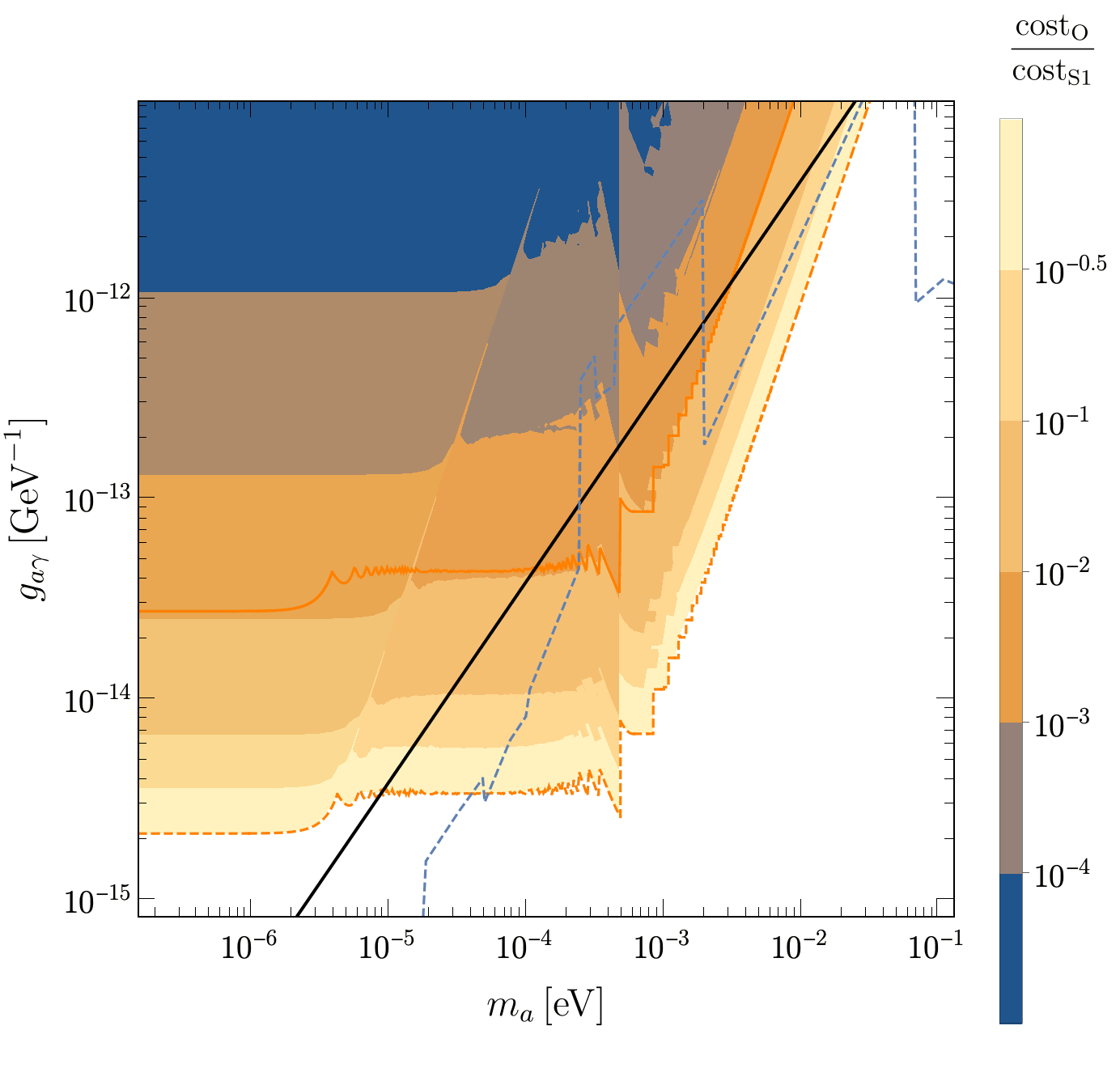}
    \caption{Contour plot of the O-type \hyperlsw cost and $a < \SI{1.3}{\m}$ relative to the cost of the corresponding S1~setup. The KSVZ model (solid black line) and and projected haloscope sensitivity (dashed light blue line) are also shown (see \cref{fig:exp} for references).}
    \label{fig:cost}
\end{figure}

Note that the choice of setup plays a crucial role in reducing costs by allowing shorter experiments.
Consider the contour plot in \cref{fig:cost}, which shows the cost of O-type optical setups and $a < \SI{1.3}{\m}$ relative to the cost for the S1~setup at each value of \ma.
The solid orange line represents the S1 sensitivity, while the dashed orange line corresponds to the O1~sensitivity.
It is evident that the O-type optical setup offers an order of magnitude better sensitivity than the S1~setup without a significant increase in cost, while the S1~sensitivity could be reached at around 100~times lower costs, confirming the expected ``magnet-driven'' scaling.
For $\gagg \gtrsim \SI{e-14}{GeV^{-1}}$, we enter the  regime of ``tunnel-driven'' costs.

Further cost reduction is possible by using a laser with a shorter wavelength.
\Cref{eq:zopt} tells us that $z \propto \lambda^{-1} a^2$, suggesting that halving $\lambda$ would permit smaller aperture~$a$ by a factor of $\sqrt{2}$ while maintaining the same experimental length and thus yielding the same sensitivity at half the cost.\footnote{This is also suggested in Ref.~\cite{Graham:2015ouw}, but to permit an increase in the length.}
The mass reach would also slightly increase due to the reduced $q \sim \ma^2/2\omega$.

Apart from costs, other factors to consider are the time and resources required to produce potentially tens of thousands of magnets.
Past experience at the LHC has shown that, after a fairly predictable ``learning curve''~\cite{10.1109/TASC.2006.869992}, three suppliers combined could produce around 400 magnets per year~\cite{10.1109/TASC.2008.920632}.
The proposed FCC-hh experiments estimate a series production time of about six years for around \num{4700} magnets, noting that the choice of superconducting material must be considered carefully due to supply chain issues~\cite{10.1140/epjst/e2019-900087-0}.
The timescales for magnet production are of the same order of magnitude as those required for the tunnel construction.
For instance, constructing the Gotthard Base Tunnel (total length of tunnels: \SI{152}{\km}) took almost 17~years~\cite{2019_NEAT_Kommission}, while construction of the Brenner Base Tunnel (total length of tunnels: \SI{230}{\km}) is expected to take 25~years (see \cref{note:bbt}).

\subsection{Other use cases for the \hyperlsw facility}\label{sec:other_uses}

While it is premature to build a detailed physics case for \hyperlsw, we nonetheless want to provide additional motivation for the significant investment required for such a facility.

\subsubsection{Uses related and complementary to axion physics}\label{sec:other_axion}

A haloscope discovery of QCD~axions would not only reveal \ma but also the energy scale related to the Peccei--Quinn~(PQ) mechanism, set by the axion decay constant \fax, thanks to the relation~\cite{1511.02867,1812.01008}
\begin{equation}
    \ma = \SI{5.69(5)}{\micro\eV} \left( \frac{\SI{e12}{\GeV}}{\fax} \right) . \label{eq:ma0}
\end{equation}
This connection by itself could already tell us a lot about axion cosmology (see e.g.\ Refs.~\cite{1510.07633,2403.17697} for reviews), as the extremely precise \ma measurement would constrain \fax at the level of 0.9\%.
We can consider two primary cosmological scenarios for QCD~axions, one where the PQ symmetry breaks before the end of inflation and remains unbroken thereafter, and another where it breaks after inflation.\footnote{An intermediate scenario~\cite{2211.06421} or various other modifications are also possible (see Ref.~\cite[Sec.~4.1]{2203.14923} for a review).}

In the pre-inflationary scenario, the initial field value is a random variable, uniformly distributed in the canonical range of $\mtx{\theta}{ini} \sim \mathcal{U}(-\pi,\pi)$~\cite{Turner:1985si}.
There is also a contribution from isocurvature fluctuations in this scenario which, depending on the detected value of \ma, could imply fine-tuning that is worse than the strong~$CP$ problem itself \cite[e.g.][]{0911.0421}.
In the post-inflationary scenario, we consider an ensemble average of random initial field values, which is equivalent to single, fixed value of $\sqrt{\langle\mtx{\theta}{ini}^2\rangle} \approx 2.6$~\cite{Turner:1985si}, with additional contributions from topological defects.
Recent simulations of axion production from cosmic string decays are not in full agreement between each other, but suggest that these contribution will increase \afrac~\cite{1707.05566,1708.07521,1806.04677,1809.09241,1906.00967,1908.03522,2007.04990,2102.07723,2108.05368,2401.17253}.
A precise measurement of \ma could favor one of the two cosmological scenarios, and potentially allow us to estimate \afrac rather than use the limiting case of $\afrac = 1$ in \cref{sec:results} to justify construction of \hyperlsw.

Furthermore, since $\gagg \propto 1/\fax$, \hyperlsw provides insights into the nature of the newly discovered particle -- although it cannot prove that the new particle is a QCD~axion.
Using \cref{eq:gagg}, we can estimate $E/N$ and, using the QCD~axion model catalogs~\cite{2107.12378,2302.04667}, we can potentially identify only a few candidate UV models, potentially increasing our confidence in the axion's QCD nature.
A similar approach was proposed in Ref.~\cite{2101.08789} for helioscope searches, which look for axions produced in the Sun.
Previous campaigns by the Brookhaven~\cite{Lazarus:1992ry}, Sumico~\cite{hep-ex/9805026,astro-ph/0204388,0806.2230}, and CAST~\cite{hep-ex/0411033,hep-ex/0702006,0810.4482,1106.3919,1302.6283,1705.02290} Collaborations will be extended by the upcoming IAXO experiment~\cite{1401.3233,1904.09155,2010.12076}, which may measure the axion's couplings~\cite{1811.09278} or mass~\cite{1811.09290}.
An LSW measurement of \gagg and, if \ma is in the suitable mass range, a haloscope measurement of \ma may further enhance IAXO's capabilities to fit \gaee, which also depends on \fax via $\gaee \propto 1/\fax$.
Furthermore, if IAXO data prefer a dominant axion-photon coupling, we may establish evidence for KSVZ models.
However, such detailed studies are beyond the scope of this work.

\subsubsection{Use cases beyond axions}\label{sec:other_nonaxion}

The specific magnet configuration required to reach \hyperlsw's primary physics goals can be re-arranged to achieve other objectives -- similar to more generic LSW experiments.
For instance, photons and gravitons can interconvert inside a magnetic field via the (inverse) Gertsenshtein effect \cite{1961_Gertsenshtein}, making it possible for LSW experiments to detect gravitational waves (GWs)~\cite{1908.00232}.
The minimal detectable GW amplitude scales as $(BL)^{-1} a^{-1/2}$, making \hyperlsw in principle very sensitive to GWs emitted below the limit set by Big Bang Nucleosynthesis~\cite[cf.][Fig.~5]{1908.00232} -- albeit only with extremely narrow spectral range if all magnets are used.

The magnets used in \hyperlsw could also be repurposed for other (axion) experiments.
Examples are CAST used prototype magnets from the LHC~\cite{hep-ex/0411033}, and the ongoing ALPS~II experiment used modified magnets from the decommissioned HERA accelerator~\cite{2004.13441}.

The larger aperture diameter of our magnets might enable previously impossible use cases, such as magnetic resonance imaging -- although the required magnetic field homogeneity would pose significant challenges.

The infrastructure of \hyperlsw could also host other particle physics experiments.
Examples include the proposed International Linear Collider~\cite{0712.2361,1306.6353,1903.01629} or the Einstein Telescope~\cite{0810.0604,10.1088/0264-9381/27/19/194002,1012.0908}, both of which may use parts of the \hyperlsw tunnel.
Similarly, ``Hyperloop'' long-distance transport systems could potentially fit inside the tunnel, even if entirely straight tunnels are not necessarily preferred due to geological or financial considerations~\cite{10.1007/s40864-021-00149-4}.
The location of \hyperlsw could be chosen by bearing such potential future uses in mind.

\section{Conclusions}\label{sec:conclusions}

Assuming the discovery of (QCD) axions in a haloscope, we have presented a blueprint for constructing \hyperlsw, an ambitious light-shining-through-a-wall (LSW) experiment aimed at probing axion properties with a purely laboratory-based setting.
Haloscopes directly measure the axion mass \ma as well as the product $\gagg^2\eta_{a}\rholoc$ of the axion-photon coupling \gagg, the axion dark matter (DM) fraction \afrac and the local DM density $\rholoc$.
Knowledge of \ma then allows for the construction of a \hyperlsw follow-up experiment, which would directly measure \gagg.
Together, these measurements determine the axion DM density $\afrac\rholoc$.
The key findings and implications of our study are summarized as follows:

\begin{itemize}
    \item Haloscopes only measure \ma and signals $\propto \gagg^2\afrac$, necessitating an independent measurement of \gagg to establish axions as the dominant form of DM.
    \item Magnets with large aperture diameter $\gtrsim\SI{1}{\m}$ enable LSW experiments with total lengths $\gtrsim\SI{100}{\km}$.
    \item \hyperlsw can probe QCD~axions by optimizing the magnetic field configuration for a known \ma. Its most ambitious incarnations achieve KSVZ sensitivity for $\SI{2}{\ueV} \lesssim \ma \lesssim \SI{45}{\meV}$.
    \item The resonant nature of \hyperlsw at large $\ma \gtrsim \mathcal{O}(\si{\meV})$ requires a non-trivial haloscope mass resolution, as well as high levels of precision in the magnet manufacturing, arrangement, and quality control, which are realistically achievable through error control or precise magnetic field profiling.
    \item \hyperlsw setups are ambitious and costly but do not require any technological breakthroughs; any technological advancements would however further extend their reach or reduce costs.
\end{itemize}

\hyperlsw shares similarities with other large-scale physics experiments, such as the LHC -- also with regards to a ``no-lose theorem''.
While \hyperlsw may not match the versatility of other experimental facilities, its potential extensions and alternative uses mentioned in \cref{sec:other_uses} should be explored more extensively.
In fact, while the LSW designs investigated here offer favorable sensitivity across a wide mass range, we did not explore all alternatives -- LSW or otherwise -- which could potentially be cheaper or more effective to study axions, in particular for larger \ma and \gagg.
One interesting option could be a helioscope such as IAXO~\cite{1401.3233,1904.09155}, especially in the range $\ma \gtrsim \SI{10}{\meV}$ where the realization of \hyperlsw setups is more difficult.
IAXO could also benefit from knowing \ma since the coupling measurement may become more precise compared to an \textit{a priori} unknown value of \ma~\cite{1811.09278,1811.09290}.

To conclude, this study represents one of the first concrete explorations of necessary follow-up strategies in case of an axion discovery, \updated{which, to our knowledge, has not yet been investigated in detail} in the literature.

\begin{acknowledgments}
We are very grateful to Gianfranco Bertone for encouraging us to be ambitious and to push to the boundaries of experimental feasibility, thus initiating this project.
For helpful discussions, we would like to warmly thank Gianluigi Arduini, Laura Covi, Loredana Gastaldo, Erik W.\ Lentz, Axel Lindner, Edoardo Vitagliano, and the staff at the Multiphysics Division of Bilfinger Nuclear \& Energy Transition GmbH.
SH has received funding from the European Union's Horizon Europe research and innovation programme under the Marie Sk{\l}odowska-Curie grant agreement No~101065579. GL is grateful to Bilfinger Nuclear \& Energy Transition GmbH for hospitality during the last stages of this work. GL and JJ are supported by the European Union’s Horizon 2020 Europe research and innovation programme under the Marie Sk{\l}odowska-Curie grant agreement No 860881-HIDDeN.
This article is based upon work from COST Action COSMIC WISPers CA21106, supported by COST (European Cooperation in Science and Technology).
\end{acknowledgments}

\appendix

\section{Further details on the form factor \textit{F}}\label{sec:Fdetails}

In this appendix we take a closer look at the maxima of the form factor~$|F|$ in \cref{eq:generalF}, which can be approximately identified with the poles of $\tan(\cdot)$~\cite{1009.4875}.
This is helpful as numerical evaluation of $F$ is challenging due to the poles and zeros of the trigonometric functions.
We therefore have to examine the region around the maxima, particularly to estimate their width compared to the precision available from haloscopes.

\subsection{Limits and (approximate) maxima}\label{sec:Fmaxima}

Consider the fully-aligned setup ($\ngr = 1$), where
\begin{equation}
    F_{N,1}(x) = \frac{\sinc(x)}{N} \, \frac{\sin(\nm y)}{\sin(y)} , \label{eq:FparN1}
\end{equation}
and where $y \equiv x \, (1+\delta)$.
The maxima of \cref{eq:FparN1} correspond to the zeros of $\sin(y)$,
\begin{equation}
    x_k = \frac{k\,\pi}{1 + \delta} \quad \text{for $k \in \mathbb{N}_0$} . \label{eq:sin_zeros}
\end{equation}

At these maxima, the form factor reduces to that of a single magnet, $F_{N,1}(x_k) = F_{1,1}(x_k) = \sinc(x_k)$.
The absolute maximum occurs at $x_0 = 0$, implying that this setup leads to the best sensitivity in the low-mass limit $\nm y \ll 1$, i.e.\ $\ma \ll 2\sqrt{\omega/\nm \ell \, (1 + \delta)}$, where $|F| = 1$.

For $\ngr > 1$, the maxima of \cref{eq:generalF} are close to the poles of the tangent, located at
\begin{equation}
    x_k = \frac{(2k + 1) \pi}{2\nset (1 + \delta)} \quad \text{for $k \in \mathbb{N}_0$} .\label{eq:tan_poles}
\end{equation}
Choosing $k = 0$ again leads to the largest form factor.
Substituting $x = q \ell/2$ with $q \approx \ma^2 / 4 \omega$, the maximum is found at
\begin{equation}
    \ma = \sqrt{\frac{2\pi\omega}{\nset\ell\,(1+\delta)}} , \label{eq:mmax}
\end{equation}
or, equivalently, for the gap parameter
\begin{equation}
    \delta = \frac{2\pi\omega}{\nset\ell\ma^2} - 1 . \label{eq:dmax}
\end{equation}
In the limit $x \to x_0$, the form factor converges to
\begin{equation}
   |F| \to \frac{1}{\nset} \left|\sinc\left(\frac{\pi}{2\nset\,(1 + \delta)}\right) \frac{1}{\sin\left(\frac{\pi}{2\nset}\right)}\right| ,
  \label{eq:Fmax}
\end{equation}
matching the expression in Ref.~\cite[Eq.~(24)]{1009.4875} after inserting the definition of~$q$.

\begin{figure}
    \centering
    \includegraphics[width=3.375in]{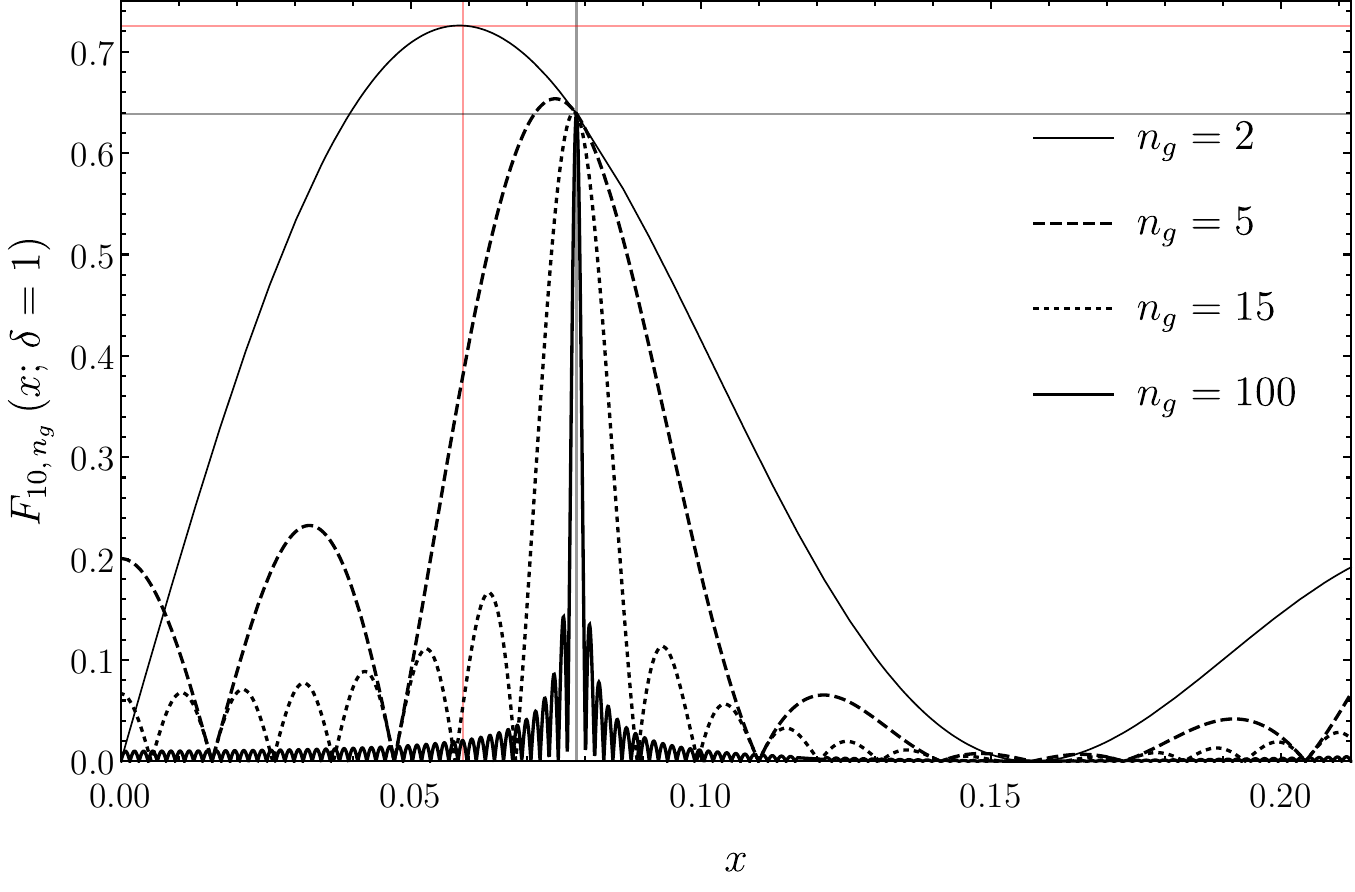}
    \caption{Form factor $|F|$ for $\nset = 10$, $\delta = 1$, and various \ngr values. The black lines corresponds to the $k = 0$ pole of the tangent, cf.\ \cref{eq:tan_poles,eq:Fmax}, while the red lines are defined by \cref{eq:x2,eq:f2}.}
    \label{fig:an_appr}
\end{figure}

The maximum of $|F|$ approaches $x_0$ as $\ngr$ increases, as illustrated in \cref{fig:an_appr} for $\nset = 10$, $\delta = 1$, and various values of \ngr.
The largest discrepancy between $|F(x_0)|$ (gray lines) and the true maximum is found for $\ngr = 2$ (red lines).
As $\ngr$ increases, the discrepancy diminishes.
For the example in \cref{fig:an_appr}, we confirmed numerically that the discrepancy is less than 2\% for the location of the maximum and less than 1\% for the value of the form factor when $\ngr > 8$.
The shift with respect to the true location of the maximum and its value is almost independent of \nset and \rgap and decreases slightly for larger values.

For $\ngr = 2$, where the discrepancy between $x_0$ and the true maximum is largest,
we can refine our estimate by observing that, for $\nset > 1$, the location and value of the absolute $|F|$ maximum are well approximated by
\begin{align}
    \tilde{x}_0 &= \frac{3}{4} \frac{\pi}{2\nset\,(1+\delta)} , \label{eq:x2} \\
    |F_{2,\nset}(\tilde{x}_0)| &= \frac{2 (2+\sqrt{2})}{3\pi} \left|\frac{(1+\delta)\,\sin\left(\frac{3\pi}{8\nset \, (1+\delta)}\right)}{\sin\left(3\pi/8y\right)}\,\right| . \label{eq:f2}
\end{align}
The red lines in \cref{fig:an_appr} correspond to \cref{eq:x2,eq:f2}, improving the estimate for the maximum from \cref{eq:tan_poles,eq:Fmax} (gray lines).

The condition in \cref{eq:mmax} implies that the largest accessible mass is $\ma = \sqrt{2\pi\omega/[\ell (1+\mtx{\delta}{min})]}$, where $\mtx{\delta}{min}$ is the minimum gap size.
To probe larger masses, the fully-alternating setup ($\nset = 1$) has to be used, with gap sizes matching the local maxima of $|F|$, located at $x_k$ in \cref{eq:tan_poles}.
With $q$ given by \cref{eq:momentum_transfer},\footnote{When requiring high accuracy,  the approximation $q \approx \ma^2/2\omega$ should be avoided for $\ma \gtrsim \SI{10}{\meV}$.} the corresponding gap sizes are
\begin{equation}
    \delta_k = \frac{(2k+1)\,\pi}{q\ell} - 1 . \label{eq:dmaxk}
\end{equation}
The local maxima for $|F|$ converge to 
\begin{equation}
    |F_{1,\nm}(x;\,\delta_k)| \to \left|\sinc(x_k)\right| \quad (x \to x_k), \label{eq:fmaxsing}
\end{equation}
showing that, in the fully-alternating configuration of \nm magnets with gap size $\delta_k$, the maxima of the form factor correspond to the value of the form factor of a single magnet of length~$\ell$.

\subsection{Expansion around the poles \textit{x\textsubscript{k}}}\label{sec:Fexpansion}

To assess if the haloscope mass determination is precise enough to realize \hyperlsw, we need to estimate the width of the form factor peaks around the poles~$x_k$.
This is especially crucial for fully-alternating setups that probe large masses due their resonant nature.
A second-order expansion in $\xi_k \equiv x - x_k$ is sufficient for narrow peaks:
\begin{align}
    \left|\frac{F_{1,\nm}(x)}{F_{1,\nm}(x_k)}\right| &= 1 + a_{N,1}^{(1)} \, \xi_k + a_{N,1}^{(2)} \, \xi_k^2 + \mathcal{O}(\xi_k^3) , \\
    a_{N,1}^{(1)} & = \cot (x_k) - \frac{1}{x_k}, \\
    a_{N,1}^{(2)} & = -\frac13 (N^2 - 1) (1 + \delta)^2 - 1 - \frac{2 a_{N,1}^{(1)}}{x_k} ,
\end{align}
with $\cot(x) \equiv 1/\tan(x)$ and where, unlike in \cref{eq:generalF}, the result is same for even and odd $\ngr = N$.\footnote{Evaluating the second derivatives requires multiple applications of {L'H\^opital}'s rule.}

Solving for the full width at half maximum, defined via $|F(x)/F(x_k)| = 1/2$, gives solutions $x_\pm$.
The haloscope's mass sensitivity can then be compared to the \emph{half width} at half maximum:
\begin{equation}
    \frac{\Delta \ma}{\ma} \approx \frac{\Delta q}{2 q} = \frac{\Delta x}{2 x} \approx \frac{x_+ - x_-}{4 x_k} .
\end{equation}
\begin{widetext}
We find that
\begin{equation}
    \frac{x_+ - x_-}{4 x_k} = \frac{\sqrt{3}}{4} \frac{\sqrt{\left[ 2(N^2 - 1) (\delta + 1)^2 + 3 (2 + \cot^2(x_k)) \right] x_k^2 + 6 \cot(x_k) \, x_k - 9}}{\left[(N^2 - 1) (\delta + 1)^2 + 3\right] x_k^2 + 6 \cot(x_k) \, x_k - 6} \simeq \frac{\sqrt{6}}{4 x_k (1+\delta)} \frac1N  \quad (N \to \infty) .\label{eq:peak_width}
\end{equation}
\end{widetext}

The width of in \cref{eq:peak_width} thus decreases as \nm increases.
As further discussed in \cref{sec:opt:m1} and \cref{sec:opt:m2}, this may imply fairly stringent requirements on the mass resolution of the preceding haloscope measurement.
In particular, at higher masses $\gtrsim \text{few}\,\text{meV}$, this may be a non-trivial prerequisite for building \hyperlsw.

\section{Optimization algorithm for \hyperlsw}\label{sec:generalstrategy}

\begin{figure}
    \centering
    \includegraphics[width=3.375in]{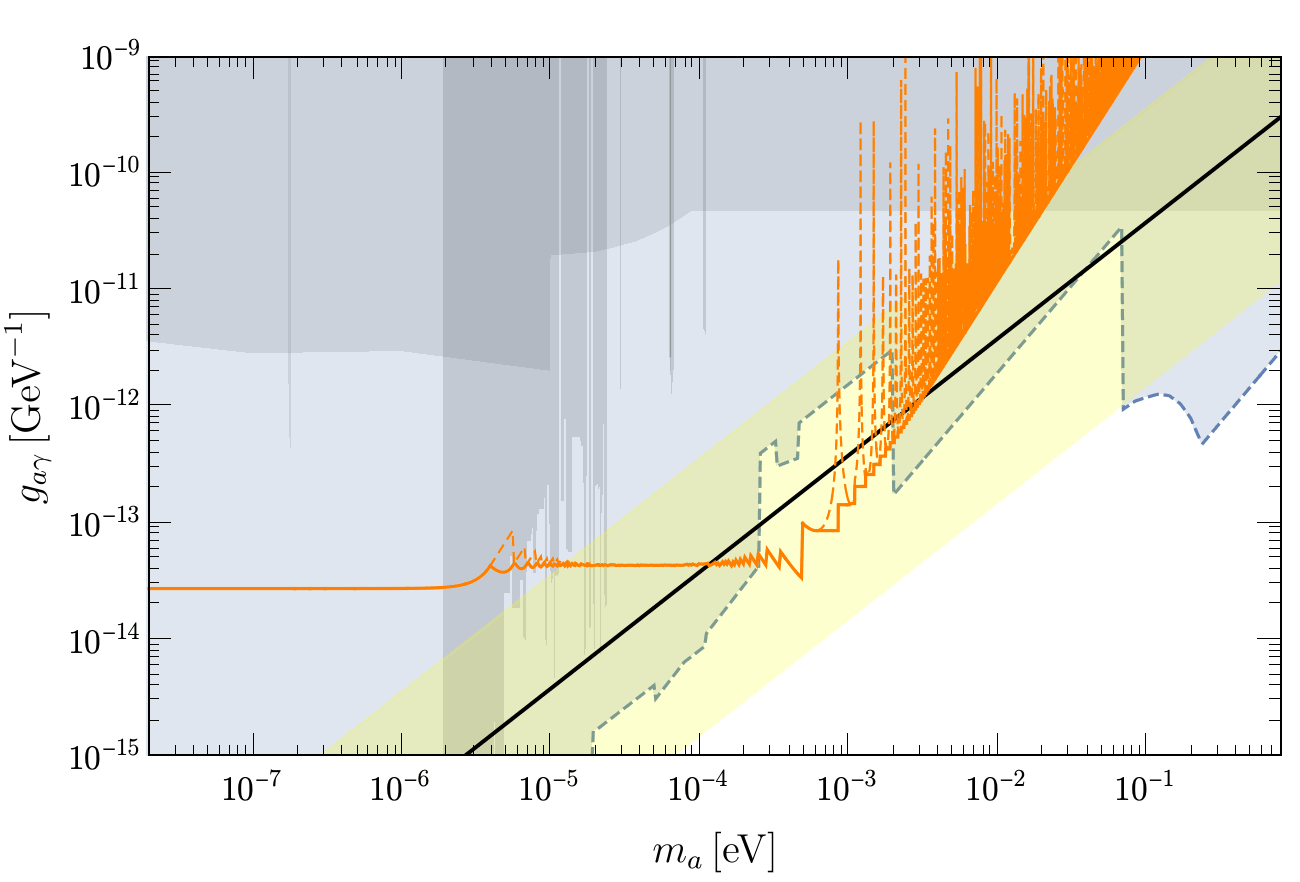}
    \caption{Sensitivity for the S1~setup. The dashed line corresponds to the sensitivity for the $\mtx{\vc{\theta}}{opt}$ parameters described in \cref{sec:generalstrategy}. The QCD axion band (yellow region), KSVZ model (solid black line), various constraints (gray region), and projected haloscope sensitivity (light blue region, dashed line) are also shown (see \cref{fig:exp} for references).}
\label{fig:opt_proc}
\end{figure}

The optimal parameter values for $\vc{\theta} = (\ngr, \, \nset, \, \delta)$, denoted by $\mtx{\vc{\theta}}{opt} = (\mtx{n}{g,opt}, \, \mtx{n}{s,opt}, \, \mtx{\delta}{opt})$, can be found by numerically maximizing the expected signal $\mathcal{S} \propto \gagg^4$ for a given value of \ma.
Such mixed-integer optimization problems are typically challenging.
Fortunately, as shown in \cref{sec:Fdetails}, the poles $x_k$ are very close to the true location of the maxima of $|F|$ -- especially for large $\nm = \ngr \times \nset$.
Here we detail how to exploit this via an algorithmic approach, with results for the S1 setup (see \cref{tab:bench_par}) shown in \cref{fig:opt_proc}.

\paragraph*{Intermediate-mass region.} The values for $\mtx{\vc{\theta}}{opt}$ are determined by choosing the longest total length while avoiding clipping losses, i.e.\ $\nm \ell \sim \zopt$.
In the ``intermediate-mass region'' ($\pi/2\nm < y < \pi/2$), we find the smallest value of the optimal $\mtx{\delta}{opt} \geq \mtx{\delta}{min}$ for a fixed value of \ma from \cref{eq:dmax} by varying \nset.
This gives us values for $(\mtx{\delta}{opt},\mtx{n}{s,opt})$.
We then compute $\mtx{n}{g,opt}$ as $\mtx{n}{g,opt} = \lfloor \zopt/\mtx{n}{s,opt} \ell (1 + \mtx{\delta}{opt})\rfloor$, where $\lfloor \cdot \rfloor$ indicates the floor function.
In other words: we find the largest number of groups \ngr with \nset magnets per group fitting in $z_{\rm opt}$.
The resulting setup is shown as dashed orange lines in the intermediate-mass region in \cref{fig:opt_proc}.
For masses with the same $\mtx{n}{s,opt}$, the best sensitivity is achieved for $\mtx{\delta}{opt} \to \mtx{\delta}{min}$ since this leads to larger $\mtx{n}{g,opt}$.
This explains the spikes observable at the end of the intermediate-mass region, $y \sim 1$, which are unavoidable due to fine-tuning effects: the form factor is highly oscillatory and spiky, and small changes in \ngr, \nset, and \rgap thus lead to a huge loss in sensitivity.

\paragraph*{Further improvements.} The above procedure is not entirely effective at the beginning of the intermediate-mass region, where $\mtx{n}{g,opt} \sim 1$.
Indeed, in this case the difference $\zopt - \mtx{n}{g,opt} \, \mtx{n}{s,opt} \, \ell (1+\delta) \lesssim \mtx{n}{s,opt} \, \ell (1+\mtx{\delta}{opt})$ is sizable, leading to an inefficient configuration much shorter than \zopt (cf.\ dashed orange line in the region of $\ma \approx \SIrange{5}{10}{\ueV}$).

To improve the setup, we follow an alternative procedure.
As shown in \cref{fig:an_appr}, for low values of \ngr, the full width at half maximum is larger than the one found higher values of \ngr.
Therefore, one can find a configuration $\mtx{\vc{\theta}}{opt}^\prime = (\mtx{n}{g,opt}^\prime, \, \mtx{n}{s,opt}^\prime, \, \mtx{\delta}{opt}^\prime)$ for which the form factor $|F|$ is maximized at $x_a^\prime \neq x_a \approx \ma^2\ell/4\omega$, but with better sensitivity at \ma compared to the $\mtx{\vc{\theta}}{opt}$~configuration.
Since we want to achieve a total length close to \zopt by adding more magnets, we fix $\mtx{n}{g,opt}^{\prime} = \mtx{n}{g,opt}+1$ and $\mtx{n}{s,opt}^\prime = \lfloor \zopt/\mtx{n}{g,opt}^\prime\,\ell\,(1+\mtx{\delta}{min})\rfloor$.
We then compute $\mtx{\delta}{opt}^\prime = \zopt/(\mtx{n}{g,opt}^\prime \, \mtx{n}{s,opt}^\prime \,\ell) - 1$.
Finally, for each \ma, the \hyperlsw sensitivity is defined as the lowest value of \gagg which can be probed with $\mtx{\vc{\theta}}{opt}$ and $\mtx{\vc{\theta}}{opt}^\prime$ configurations.
We show the outcome of this procedure as a solid line in the intermediate-mass region of \cref{fig:opt_proc}.
These differ from the $\mtx{\vc{\theta}}{opt}$ parameters (dashed lines) only at the beginning of the intermediate-mass region, implying that $\mtx{\vc{\theta}}{opt}^\prime$ leads to a better sensitivity for small \ngr, while $\mtx{\vc{\theta}}{opt}$ is better at larger \ngr.

\paragraph*{High-mass region.} For higher masses, $y > \pi/2$, we slightly modify the intermediate-mass-region procedure for $\mtx{\vc{\theta}}{opt}$ by finding the lowest value of $k$ for which $\delta_k \geq \mtx{\delta}{min}$ in \cref{eq:dmaxk}.
For this value of $\delta_k$, the form factor reduces to the single-magnet form factor $F_{1,1}$ as shown in \cref{eq:fmaxsing}, and the sensitivity is given by the dashed orange line in \cref{fig:opt_proc} at masses $\ma \gtrsim \SI{1}{\meV}$.
To avoid the zeros of $F_{1,1}$, we propose to shorten the length of the single magnet, $\ell \mapsto \ell^\prime$ (see \cref{sec:opt:m2plus}).
The resulting sensitivity is shown as a solid orange line in \cref{fig:opt_proc}.
In this case, $\ell^\prime$ is chosen to maximize $|F_{1,1}|$, whose maxima are given by \cref{eq:max_sinc}, while fulfilling the condition in \cref{eq:tangent_poles}.

A word of caution for masses corresponding to $x \lesssim 4.49$, which is the location of the second local maximum of $|\sinc(x)|$ (the first maximum is at $x = 0$).
For these masses, the previous maximum would correspond to $\ell^\prime = 0$, and we need to employ a different strategy.
The high-mass regime starts at $m_{a,\text{crit}} = \sqrt{2\pi\omega/\ell (1 + \mtx{\delta}{min})}$, while the best sensitivity is reached at $\ma^\ast = \sqrt{2\pi\omega/\ell}$, due to the product $L^4\,|F_{1,1}|^4$ in the computation of $\pag^2$ in \cref{eq:p_agamma}.
Therefore, for $x \lesssim 4.49$, for masses $m_{a,\text{crit}} < \ma < \ma^\ast$ shortening the magnet length would not give any improvement, while for larger masses one can improve the sensitivity by choosing a length $\ell^\prime = q(\ma^\ast)\,\ell / q(\ma)$.
This leads to the sensitivity shown as a solid orange line in \cref{fig:opt_proc} for $\ma \gtrsim \SI{0.5}{\meV}$.

\begin{figure}
    \centering
    \includegraphics[width=3.375in]{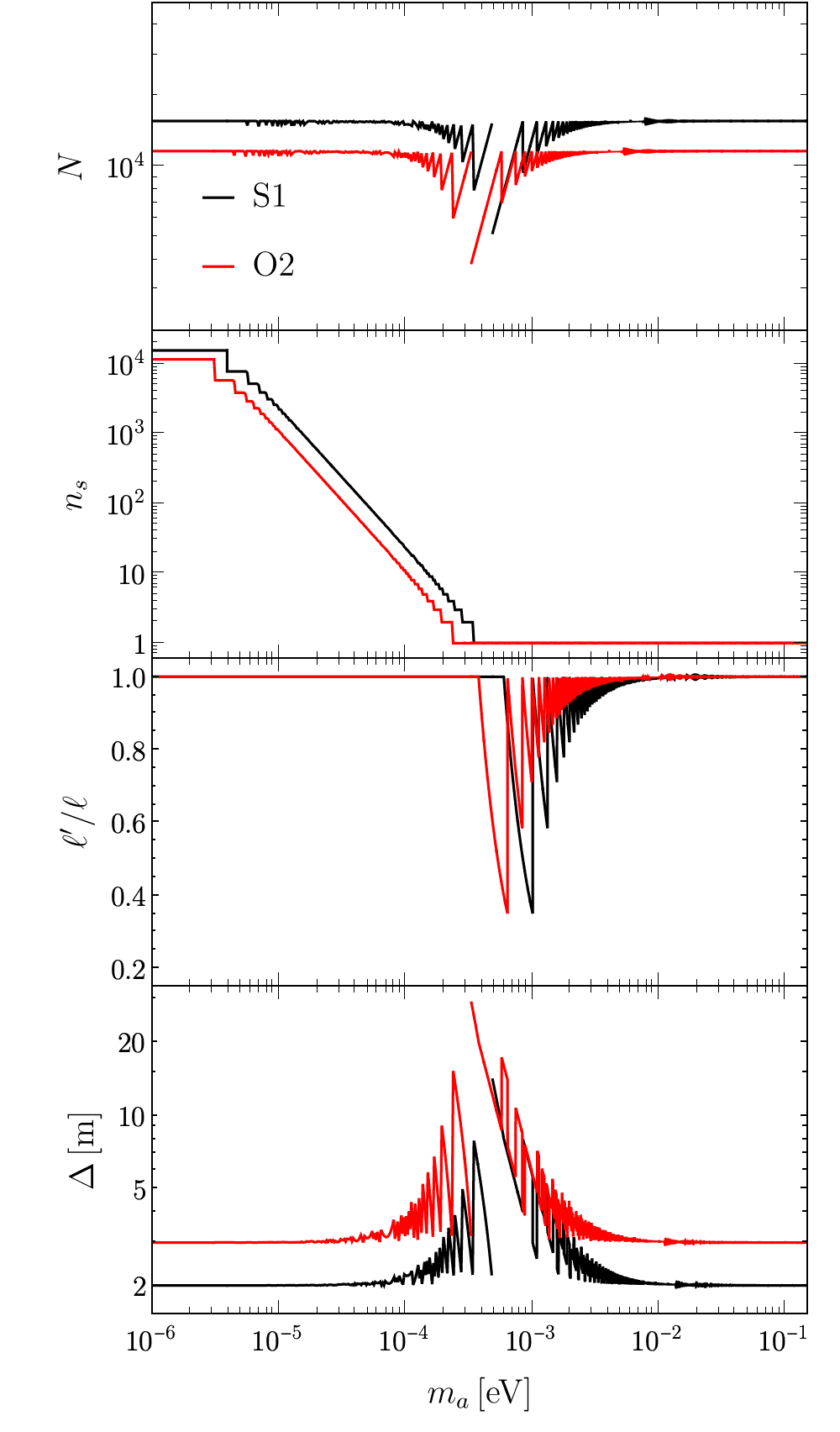}
    \caption{Optimal parameter choices as a function of $\ma$ for S1 (black) and O2 (red) setups from \cref{tab:bench_par}.\label{fig:M1_par}}
\end{figure}

Following the procedure described above, we obtain the optimal values for \nset, \ngr, $\delta$, and $\ell^\prime$ (where $\ell^\prime \neq \ell$ only in the high-mass limit) for each mass \ma to measure the coupling \gagg with the desired precision of $\Pi = 2\%$.
We show the values of the optimal parameters for the S1 (black line) and O2 (red line) setups in \cref{fig:M1_par}.

\section{Experimental components}\label{sec:technology}

For reference, we summarize some of the available technology to inform our choices for the \hyperlsw benchmark setups and add a number of technical details regarding the clipping losses and filling the experimental setup with a buffer gas.

\subsection{Magnets}\label{sec:magnets}

In \cref{table:magnets}, we list a number of available or proposed magnet designs to extend the selection of magnets available at the time of publication of Ref.~\cite{1009.4875}.
We list the magnetic field strength~$B$, aperture diameter~$a$, and length~$\ell$.
In the main text, we choose a magnet that closely resembles the proposed MADMAX magnet due to its large aperture and sizable magnetic field.

\begin{figure}
    \centering
    \includegraphics[width=3.375in]{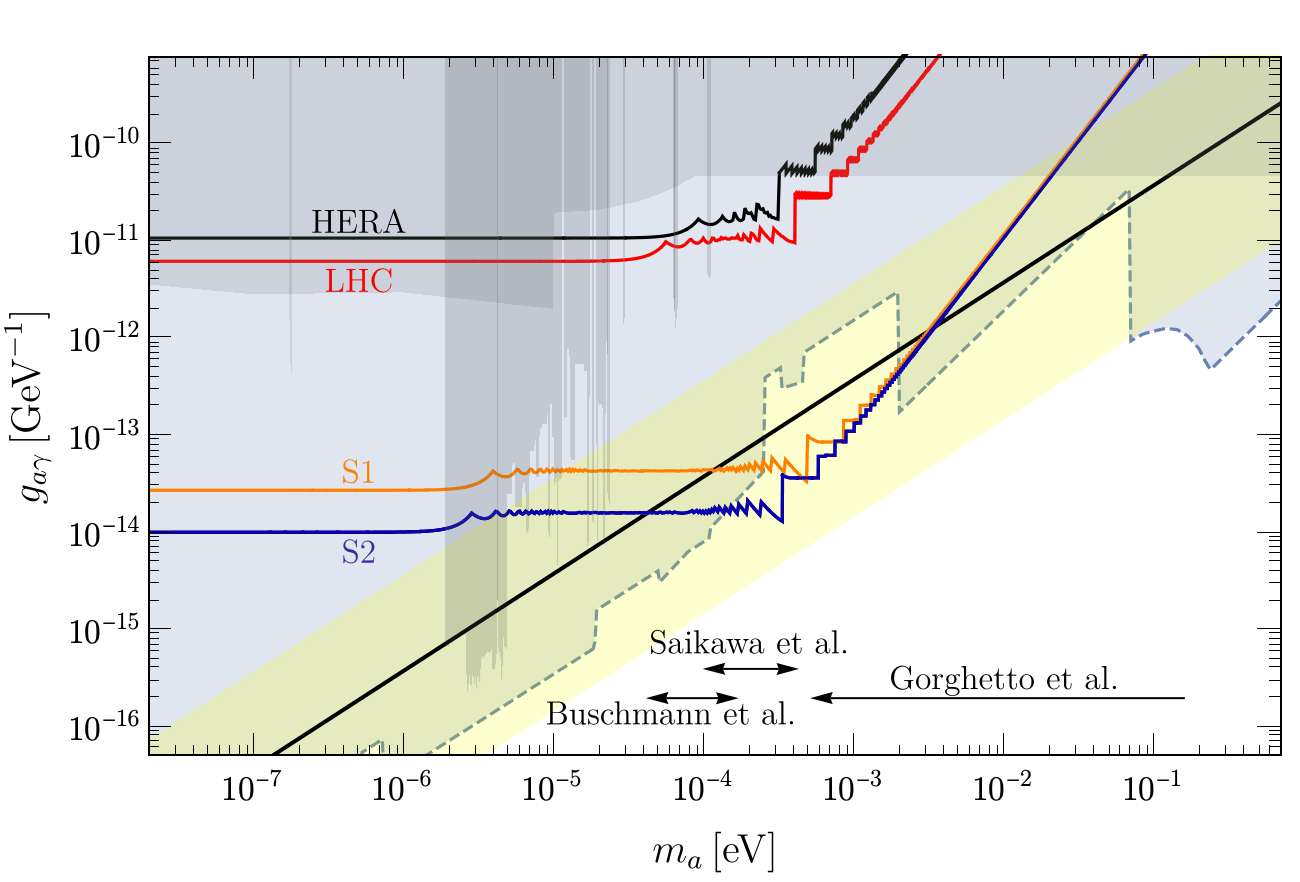}
    \caption{Sensitivity for LHC (black) and HERA (red) magnets from \cref{table:magnets}, consistently compared to the S1 (orange) and S2 (blue) setups from \cref{tab:bench_par}. The QCD axion band (yellow region), KSVZ model (solid black line), various constraints (gray region), and projected haloscope sensitivity (light blue region, dashed line) are also shown (see \cref{fig:exp} for references).}
\label{fig:sens_LHC}
\end{figure}

In \cref{fig:sens_LHC}, we show the sensitivity (as defined in the main text) for various other magnets listed in \cref{table:magnets}, using  $\mtx{\Delta}{min} = 0$ for the HERA and LHC magnets and otherwise the parameters for the ``S~setup'' from \cref{tab:bench_par}.
Clearly, even without a minimal gap, HERA and LHC magnets cannot probe the QCD~axion band, which is due to their small aperture.

\begin{table}[t!]
\caption{Overview of proposed or used magnets in other experiments (see also Refs.~\cite{10.1088/0034-4885/54/5/001} and \cite[Sec.~32]{10.1093/ptep/ptac097}). The value for the aperture diameter $a$ quoted for HERA (ALPS~II) magnet corresponds to the lowest among these magnets.\label{table:magnets}}
\smallskip
\begin{tabular}{lS[table-format=2.1]S[table-format=1.2]S[table-format=2.1]c}
\toprule
    Magnet & \multicolumn{1}{c}{$B$ [\si{\tesla}]} & \multicolumn{1}{c}{$a$ [\si{\m}]} & \multicolumn{1}{c}{$\ell$ [\si{\m}]} & Reference(s) \\
\midrule
    CMS & 3.8 & 6.30 & 12.5 & \cite{10.1109/TASC.2023.3336838} \\
    FCC-ee (IDEA) & 2.0 & 2.10 & 6.0 & \cite{10.1140/epjst/e2019-900045-4} \\ 
    FCC-hh (MD) & 16.0 & 0.05 & 15.8 & \cite{10.1140/epjst/e2019-900087-0} \\
    HERA & 4.7 & 0.07 & 8.8 & \cite{10.2221/jcsj.31.360} \\ 
    HERA (ALPS~II) & 5.3 & 0.09 & 8.8 & \cite{2004.13441} \\ 
    LHC & 8.3  & 0.06 & 14.3 & \cite{10.5170/CERN-2004-003-V-1} \\ 
    MADMAX & 9.0 & 1.35 & 6.0 & \cite{10.1109/TASC.2020.2989478,10.1109/TASC.2023.3315201} \\
    TEVATRON & 4.4 & 0.08 & 6.4 & \cite{10.1146/annurev.ns.35.120185.003133} \\ 
\bottomrule
\end{tabular}
\end{table}

\subsection{Optics and clipping losses}\label{sec:optics}

As explained in Ref.~\cite{1009.4875}, we have to take into account clipping losses, which effectively limit the total length of the experiment and define its optimal length.
To see this, consider the spot size $w(z)$ of a Gaussian beam at distance $z \in [0, z_\nm]$ with waist size~$w_0$, where $z_\nm \approx \nm\ell(1+\delta)$ is the length of one part of the experiment:
\begin{equation}
    \frac{w(z)}{w_0} = \sqrt{1 + \left(\frac{z-z_\nm}{\mtx{z}{R}}\right)^2} ,
\end{equation}
with the Rayleigh length $\mtx{z}{R} = \pi w_0^2 \nref / \lambda$.
Assuming that the boost factor $\beta$ is the same in both the generation and regeneration regions ($\mtx{\beta}{g} = \mtx{\beta}{r} = \beta$), the sensitivity of the experiment approximately scales with $\beta^{-1/2} L^{-1}$, with $\beta$ given in \cref{eq:boost}.
Its maximum for $\nref = 1$ is implicitly given by~\cite{1009.4875}
\begin{equation}
    \ee^{-\zeta} \left(\frac{\zeta}{2} - 1\right) - \beta_0^{-1} = 0 , \label{eq:z_opt_implicit}
\end{equation}
where $\zeta \equiv \pi a^2/4\lambda z$, $a$ is the aperture (diameter) of the magnets.
The quantity $\beta_0^{-1} \equiv -\sum_{i} \ln (R_i)$ depends on the coefficients $R_i$, which are reflectivity coefficients related to other losses (mirror transmissivity and imperfections, round trips, etc.).\footnote{Note that the authors of Ref.~\cite{1009.4875} define the magnet aperture via its radius, while we use its diameter, implying that \cref{eq:z_opt_implicit} reduces to Eq.~(32) in \cite{1009.4875} after replacing $a\to 2\,a$.}

\updated{On the one hand, the approximate solution in \cref{eq:zopt} can be obtained by ignoring the term in brackets in \cref{eq:z_opt_implicit}~\cite{1009.4875}.}
\updated{On the other hand, the solution can be found numerically. To guarantee the successful convergence of the root finding algorithm, we need to bracket the correct solution, which only exists} when $\beta_0 > 2\ee^3 \approx \num{40}$.
Since \cref{eq:z_opt_implicit} has a maximum at $\zeta = 3$ and, for $\beta_0^{-1} = 0$, a zero at $\zeta = 2$, there should then be a zero \updated{in the interval} $\zeta \in [2,\,3]$.
Moreover, for $\zeta \gg 0$, we can approximate \cref{eq:z_opt_implicit} as $\zeta\ee^{-\zeta}/2 - \beta_0^{-1}$, which has zeros at $-W_0(-2\beta_0^{-1})$ and $-W_{-1}(-2\beta_0^{-1})$, where $W_k$ denotes the $k$-branch of the Lambert~$W$ function.
Since $\beta_0 \gtrsim 40$ implies that $-W_0(-2\beta_0^{-1}) \lesssim 0.06$ and $-W_{-1}(-2\beta_0^{-1}) \gtrsim 4.5$, the latter is the only valid solution in the interval $3 < \zeta < -W_{-1}(-2\beta_0^{-1})$, which also leads to the longer experiment.

\updated{Choosing} $z = \zopt$ gives \updated{a total} boost factor that is smaller than \updated{the ``intrinsic'' boost}~$\beta_0$.
\updated{For} the example in \cref{eq:zopt}, we find $\beta \approx 0.85\,\beta_0$.
Furthermore, \cref{eq:z_opt_implicit} tells us that larger $\beta_0$ leads to lower values of \zopt.
This justifies our \updated{simplification} for O-type setups in the main text, where we assumed that the total length of the experiment is \updated{$2 \times \zopt$, where \zopt is calculated for} the regeneration \updated{part of the experiment} (ignoring that $\mtx{\beta}{r} > \mtx{\beta}{g}$).
While the regeneration boost factor is larger, the corresponding cavity contains at most a small number of photons, drastically reducing heating issues of the mirrors.

\subsection{Buffer gas filling}\label{sec:gas_filling}

Filling the LSW setup with a gas changes the refractive index \nref inside the magnetic fields, allowing us to achieve values of $\nref > 1$.
By adjusting \nref we can, in principle, change the momentum transfer $q$ and increase the sensitivity to some \ma values via \cref{eq:tangent_poles}.
The feasibility of this approach has already been demonstrated in a number of LSW experiments~\cite{0712.3362,1004.1313,1110.0774}.

A non-optimal form factor~$|F(x)|$ can be improved by shifting its argument, i.e.\ the phase factor related to the momentum transfer.
In \cref{sec:opt:m2plus} we shortened the magnet length~$\ell$ to achieve this by shifting to the \emph{previous} maximum of~$|F|$.
In contrast, a buffer gas with $\nref > 1$ \emph{increases} the phase difference between the axion and the photon, cf.\ \cref{eq:momentum_transfer}, and we can thus move to the subsequent maximum.
The required shift it
\begin{equation}
    \Delta x \sim \pi/2 \Rightarrow \Delta q \, \ell \approx (\nref - 1) \, \omega\ell \sim \pi ,
\end{equation}
which corresponds to $\nref - 1 \sim \num{1.5e-7}$ for the S1~setup.

The value of \nref can be adjusted by changing the pressure via the Lorentz--Lorenz formula (see Ref.~\cite{1302.5647})
\begin{equation}
    \nref - 1 = \left(\mtx{n}{w} - 1\right) \frac{\mtx{P}{w}}{\mtx{P}{c}}  \left(\frac{\mtx{P}{w}}{\SI{1}{\bar}}\right) ,
\end{equation}
where $\mtx{n}{w}$ and $\mtx{P}{w}$ are the refractive index and pressure at room temperature, whereas $\mtx{P}{c}$ in the pressure inside the cooled LSW setup.
For the S1~setup, the required pressure is of order $\SI{0.5}{\milli\bar}$ and, for the currently ongoing ALPS~II experiment, it has been estimated that the round trip losses in the optical cavities due to Rayleigh scattering on He~atoms at that pressure are of the order $R = \num{2e-8}$~\cite{1302.5647}.
Since a round trip in ALPS~II setup has a length of $2L \approx \SI{200}{\m}$~\cite{1302.5647}, an experiment with round trip length $2L = \SI{100}{\km}$ would have expected losses of about $\num{500}\,R = \num{e-5} \sim \beta_0^{-1}$.

In summary, filling \hyperlsw with a buffer gas would at least be challenging, as it already limits the realizable boost factor of the cavity.
This is particularly true for the envisioned O-type setup regeneration boost factor of $\mtx{\beta}{r} = \num{e6}$.

\section{Experimental errors}\label{sec:errors}

For large \ma, i.e.\ $y \gtrsim 1/\nm$, \hyperlsw improves the sensitivity of a fully-aligned setup by resonantly enhancing the form factor~$|F|$ in \cref{eq:integralF}.
This requires accurate magnet placement to achieve gaps of size $\Delta$, which we consider in \cref{sec:positioning_errors}.
Moreover, $|F|$ depends on the (consistency of the) location and shape~$f(z)$ of the magnetic field in \cref{eq:integralF}, which we consider in \cref{sec:b_field_profiles}.

\subsection{Magnet positioning errors}\label{sec:positioning_errors}

One option to arrange \hyperlsw is to measure $\Delta$ between each pair of magnets.
If each measurement has some uncertainty \siggap, the positioning error of the last magnet has a (correlated) uncertainty of size $\sqrt{\nm - 1} \, \siggap$.
Given the large number of magnets used in \hyperlsw, this can be problematic as errors accumulate.

Alternatively, one could measure the absolute position of the magnets and avoid error accumulation.
For instance, if satellite navigation systems can be used, we may achieve $\siggap = \SI{20}{\cm}$ (or better) for the Galileo High Accuracy Service \cite{10.2878/581340}.
A surveying station can further improve the satellite navigation error over the course of a few days, which has been demonstrated to give precision of $\siggap = \SI{0.7}{\cm}$ during construction of the Brenner Base tunnel \cite{2013_leica,2022_bbt}.
Moreover, other advanced surveying techniques exist that should allow absolute positioning relative to fixed points of the setup with an accuracy of $\siggap \ll \SI{1}{\cm}$~\cite{Arduini}.

We can thus reasonably suppose an absolute positioning error of $\siggap = \SI{1}{\cm}$ for each magnet~$j$, translating into relative offsets $\epsilon_j^\delta \sim \mathcal{N}(0, \sigma_\delta^2)$, where $\sigma_\delta \equiv \siggap/\ell$.
Starting from \cref{eq:integralF} and following Ref.~\cite[Eq.~(19)]{1009.4875}, we find
\begin{widetext}
\begin{align}
    F &= \frac{1}{L} \int_0^{(z_\nm '+ \epsilon_N^\delta)\ell} \! \dd z \, f(z) \, \ee^{\ii q z}
    = \frac{1}{\nm} \int_0^{z_\nm' + \epsilon_N^\delta} \! \dd z' \, f(z') \, \ee^{2\ii x z'}
    = \frac{1}{\nm} \sum_{j=1}^{\nm} \left(\int_{z_{j-1}'}^{z_j' + \epsilon_j^\delta} \! \dd z' \, f(z') \, \ee^{2\ii x z'}\right) \\
    &= \frac{1}{\nm} \sum_{j=1}^{\nm} \left[\left(\int_0^1 \! \dd z_j'' \, f_{j-1} \, \ee^{2\ii x z_j''}\right) \ee^{2\ii x (z_{j-1}' + \epsilon_j^\delta)} \right]
    = \frac{1}{\nm} \frac{\ee^{2\ii x} - 1}{2\ii x} \sum_{j=1}^{\nm} f_{j-1} \ee^{2\ii x (z_{j-1}' + \epsilon_j^\delta)} \\
    \Rightarrow |F|^2 &= \left(\frac{\sinc(x)}{\nm}\right)^2 \left| \sum_{j=0}^{\nm - 1} f_{j} \ee^{2\ii x (z_j' + \epsilon_{j+1}^\delta)} \right|^2 , \label{eq:F_with_errors}
\end{align}
\end{widetext}
where we defined $z' \equiv z/\ell$, $z_j' \equiv j (1+\delta)$, $z_j'' \equiv z' - z_j'$, and used that $f(z') = 0$ in the gaps and $f_j \in \{-1,+1\}$ inside the magnetic fields.\footnote{For relative distance measurements, replace $\epsilon_j \mapsto \sum_{k=1}^{j} \epsilon_k$.}

For each half of the experiment, we can compute the effect of positioning uncertainties on $|F|^2$, and thus on the conversion probability, with Monte Carlo (MC) simulations, drawing random values of $\epsilon_j^\delta \sim \mathcal{N}(0, \sigma_\delta^2)$.
To better understand the outcome of these computations, consider the most relevant case, the fully-alternating setup, i.e.\ $\nset = 1$ and $f_j = (-1)^j$.
For simplicity, further assume that $\ngr = \nm$ is an odd number.
Again defining $y \equiv x (1 + \delta)$, the expectation value of $|F_{1,\nm}|^2 = F_{1,\nm} \bar{F}_{1,\nm}$ is then given by
\begin{widetext}
\begin{align}
    \frac{\mathrm{E}\left[|F_{1,\nm}|^2\right]}{\sinc^2(x)} &= \frac{1}{\nm^2}\prod_{i=1}^{\nm} \int \! \dd \epsilon_i^\delta \, \frac{\ee^{-(\epsilon_i^\delta)^2/2\sigma_\delta^2}}{\sqrt{2\pi}\sigma_\delta} \sum_{k,j=0}^{\nm-1} f_j \, f_k \, \ee^{2\ii x [(j-k)(1+\delta) + \epsilon_{j+1}^\delta - \epsilon_{k+1}^\delta]} \\
    &= \frac{1}{\nm^2}\left(\sum_{k=0}^{\nm-1} f_k^2 + \sum_{k,j \neq k} f_k f_j \, \ee^{2\ii x (j-k)(1+\delta) - 4 x^2 \sigma_\delta^2}\right)
    = \frac{1}{\nm} + \frac{\ee^{- 4 x^2 \sigma_\delta^2}}{\nm^2} \sum_{k,j \neq k} (-1)^{j+k} \, \ee^{2\ii x (j-k)(1+\delta)} \\
    &= \frac{1}{\nm} + \frac{\cos^2(\nm y) - \nm \cos^2(y)}{\nm^2\cos^2(y)} \, \ee^{- 4 x^2 \sigma_\delta^2} \to \frac{1}{\nm} + \frac{\nm-1}{\nm} \, \ee^{- 4 x_k^2 \sigma_\delta^2} \quad (x \to x_k) . \label{eq:placement_uncert_shift}
\end{align}
\end{widetext}

\Cref{eq:placement_uncert_shift} shows that the random placement errors will introduce a systematic shift in the estimate, which becomes more significant at larger~$x$.
There are also statistical fluctuations, but we do not perform the related, rather involved computation of $\mathrm{Var}[|F_{1,\nm}|^2]$.
Also note that $\mathrm{E}\left[|F_{1,\nm}|^2\right] \to \sinc^2(x) \cos^2(\nm y) / \nm^2 \cos^2(y)$ for $\sigma_\delta \to 0$, in agreement with \cref{eq:generalF}.

\begin{figure*}
    \centering
    \includegraphics[width=6.8in]{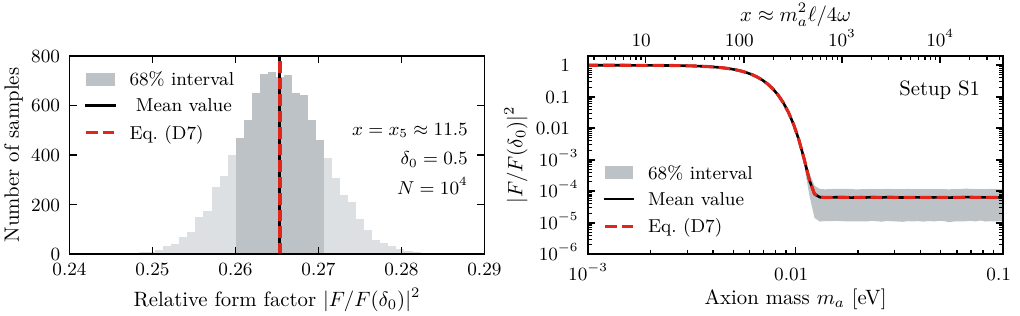}
    \caption{Form factor uncertainties from positioning errors, based on \num{10000} simulations and $\siggap = \SI{1}{\cm}$. We show an example distribution for $x = x_5$, $\Delta = \SI{2}{\m}$, and $\ell = \SI{4}{\m}$ (left panel) and the uncertainties for the optimized setup~S1 across different masses (right panel).}
    \label{fig:placement_uncert}
\end{figure*}
\Cref{fig:placement_uncert} shows an example for the ensuing distribution for $x = x_5 \approx 11.5$ (left panel) and the effect of uncertainties for the optimal parameters in our proposed setup~S1 across the entire \ma range (right panel).
In both cases, we assume that $\siggap = \SI{1}{\cm}$ and perform \num{10000} MC simulations for each configuration.

In particular, for values of $\ma \gtrsim \SI{4}{\meV}$, the positioning uncertainties cause a significant shift in the expectation value of more than 10\% (plus sizable scatter) in the form factor, and thus in the expected number of counts.
This is not entirely surprising since only the high-mass region strongly depends on resonant enhancement, and thus requires an accurately positioned experimental setup.
In any case, we can compute the impact of the positioning uncertainties and, even for a conservative value of $\siggap = \SI{1}{\cm}$, they would only affect the multi-meV mass region.

\subsection{Realistic magnetic field profiles}\label{sec:b_field_profiles}

\hyperlsw likely requires thousands of magnets, and we thus have to assume that the magnet profiles $f(z)$ will be subject to production errors such as asymmetries and length variations.
Moreover, $f(z)$ will not have the top-hat profile that we assume in the main text, but smoothly decrease outside of the nominal ``iron length''~$\ell$.

There is a variety of functions to describe such a behaviour and, for computational efficiency, we choose the ``smooth(er) step'' functions $\mathrm{smst}(z;a_1,a_2)$~\cite{2003_Perlin}, which define a step between $a_1 < a_2$ such that the function vanishes for $z < a_1$ and is unity for $z > a_2$.
This allows us to construct smooth top hat between $-\zeta < z < 1  + \epsilon^\ell+\zeta$, where $\zeta > 0$ defines the the size of the smooth step and $\epsilon^\ell$ is a small shift in the magnetic field length (both in units of~$\ell$).
The magnetic form factors $f_j(z')$ are then given by
\begin{widetext}
\begin{align}
    f_j(z') &= f_j \; \mathrm{smst}(z'+\epsilon^\zeta; -\zeta, 0) \; \mathrm{smst}(-z'-\epsilon^\zeta; -1-\epsilon^\ell-\zeta, -1-\epsilon^\ell) ,\label{eq:bfield_profile_smooth} \\
    \text{with} \quad \mathrm{smst}(z;a_1,a_2) &= \frac{ (z - a_1)^3 (6 z^2 + 3 z a_1 + a_1^2 - 5 (3 z + a_1) a_2 + 10 a_2^2)}{(a_2-a_1)^5} \, \Theta(z - a_1) \, \Theta(a_2 - z) + \Theta(z - a_2) , \label{eq:smoother_step}
\end{align}
where $f_j \in \{-1,+1\}$ and where $\epsilon^\zeta$ is a small offset (in units of~$\ell$) from the magnetic field center.
We can then follow the derivation in \cref{sec:positioning_errors}, defining $z_j'' = z' - z_j'$ with $z_j' = j(1+\delta) + \sum_{k=1}^{j} (\epsilon_k^\ell + \epsilon_k^\zeta)$ and $\tilde{\ell}_j \equiv 1 + \epsilon_j^\ell + \zeta$ to find
\begin{align}
    F = \frac{1}{\nm} \int_{-\zeta}^{z_\nm' + \zeta + \sum_k\epsilon_k^\ell} \! \dd z' \, f(z') \, \ee^{2\ii x z'}
    = \frac{1}{\nm} \sum_{j=1}^{\nm} \left[\left(\int_{-\zeta}^{1+\epsilon_j^\ell+\zeta} \! \dd z_j'' \, f(z_j'') \, \ee^{2\ii x z_j''}\right) \ee^{2\ii x z_{j-1}'} \right] ,
\end{align}
\begin{align}
    \text{where} \quad \int_{-\zeta}^{1+\epsilon_j^\ell+\zeta} \! \dd z'' \, f(z'') \, \ee^{2\ii x z''} &= f_j \, \frac{15 (3- \zeta^2 x^2) \, \sinc(\zeta x) - 3 \cos(\zeta x)}{\zeta^4x^4} \; \tilde{\ell}_j \, \sinc(\tilde{\ell}_j x) \, \ee^{\ii x (1 + \epsilon_j^\ell)} \\
    &\simeq f_j \,\tilde{\ell}_j \, \sinc(\tilde{\ell}_j x) \, \ee^{\ii x (1 + \epsilon_j^\ell)} \quad (\zeta x \to 0) .
\end{align}
\end{widetext}

\begin{figure*}
    \centering
    \includegraphics[width=6.8in]{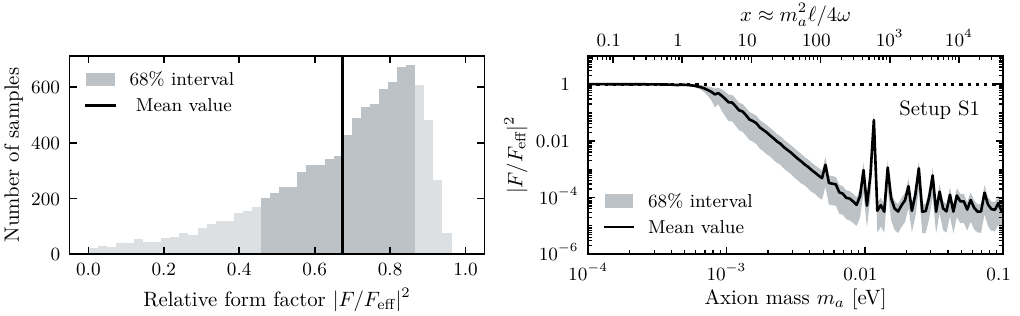}
    \caption{Effects of magnetic field profile uncertainties on the form factor, based on \num{10000} simulations and $\sigma_\ell \ell = \sigma_\zeta \ell = \SI{1}{\cm}$. We show an example distribution for $x = x_5$, $\Delta = \SI{2}{\m}$, and $\ell = \SI{4}{\m}$ (left panel) and the uncertainties for our optimized  setup~S1 across different masses (right panel).}
    \label{fig:bfield_uncert}
\end{figure*}

For $\epsilon^\zeta = 0$, we can absorb the effect of the smoother profile into a multiplicative factor, i.e.\ into an effective magnetic field strength.
For $\epsilon^\ell = 0$, and in the limit of $x \to 0$, this factor is $\sqrt{1+\zeta} > 1$, and we denote magnitude of the associated form factor with $\mtx{F}{eff}$.

Based on experience from the LHC, the systematic differences for $\epsilon^\ell$ between manufacturers was \SI{14}{\mm}, while the intrinsic scatter for each manufacturer was less than \SI{3}{\mm}~\cite{10.1109/TASC.2005.864294}.
We use these findings to define the typical uncertainty and simulate the proposed \hyperlsw configuration for setup~S1 with $\epsilon_j^\ell \sim \mathcal{N}(0,\sigma_\ell^2)$ and $\epsilon_j^\zeta \sim \mathcal{N}(0,\sigma_\zeta^2)$, where we choose $\sigma_\ell$ and $\sigma_\zeta$ to both correspond to absolute values of \SI{1}{\cm}.

To estimate $\zeta$, we use a publicly available simulation for the magnetic field profile of the upcoming IAXO experiment~\cite{1401.3233,1904.09155,2010.12076}, which is available in Ref.~\cite{10.5281/zenodo.10650613}.
The magnet has an ``iron length'' of $\ell = \SI{6.7}{\m}$ and an aperture diameter of $a = \SI{0.7}{\m}$, and we find a best-fitting value for smooth step of around $\zeta = 0.35$.
In fact, the relevant quantity that sets the size of $\zeta$ is $a$, as has been established by the expansion of fringe fields, with $\zeta = \eta_\zeta \, a/\ell$ and $\eta_\zeta \sim \numrange{4}{5}$~\cite[e.g.][Fig.~3]{10.1063/1.1718806} (see also Ref.~\cite{10.1142/p899}).
This is in line with the simulated IAXO magnet, where $\zeta = 3.5 \, a/\ell$.
If we used the same relation for the magnetic in setup~S1, we would have $\zeta = 1.1$ and, since $\delta = 0.5$, there would be a significant overlap between the stray fields of two consecutive magnets.
To simplify the computation, we choose $\zeta = 0.2$.

We again perform \num{10000} MC simulations for each configuration of the optimized S1~setup, assuming the uncertainties listed above.
We also again include an example distribution as the left panel of \cref{fig:bfield_uncert}, which shows that the accumulating, correlated errors, although small, can lead to highly asymmetric distributions of $|F|^2$.
Note that we normalized the results with respect to the ideal, effective form factor $\mtx{F}{eff}$ at low \ma i.e.\ without uncertainties.

In the right panel of \cref{fig:bfield_uncert} we find both a systematic shift and scatter in the distribution of $|F|^2$, which is the same qualitatively similar to what we observed for the positioning errors in \cref{sec:positioning_errors}.
These effects are most pronounced at larger \ma, again due to the resonant enhancement, which requires an accurately arranged setup.
For the choice of uncertainties considered, the expected $|F|^2$ sees a shift of more than 10\% for $\ma \gtrsim \SI{0.6}{\meV}$.
At $\ma \gtrsim \SI{10}{\meV}$, note that the figure might show some computational issues due to numerical cancellation effects.
However, it is clear that, for those masses, the magnetic field profile uncertainties will be an issues.
Again, lower uncertainties will improve this situation, while individually measured magnetic field could allow us to ameliorate these issues by suitably arranging the magnets or gaps to compensate for the variation in magnetic field profiles.

\bibliographystyle{JHEP_mod}
\bibliography{auto,custom}

\end{document}

%% file: macros.tex
\newcommand{\emth}[1]{\ensuremath{#1}\xspace}
\newcommand{\mtx}[2]{#1_\text{#2}}
\newcommand{\newmm}[2]{\newcommand{#1}{\emth{#2}}}
\newcommand{\orcid}[1]{\xspace\href{https://orcid.org/#1}{\color[HTML]{A6CE39}\faOrcid{}}\xspace}

\newmm{\dd}{\mathrm{d}}
\newmm{\ee}{\mathrm{e}}
\newmm{\ii}{\mathrm{i}}
\newcommand{\vc}[1]{\boldsymbol{#1}}
\newmm{\sinc}{\mathrm{sinc}}
\newmm{\prob}{p}

\newmm{\alphaEM}{\alpha}

\newcommand{\ax}{a}
\newmm{\fax}{f_\ax}
\newmm{\ma}{m_\ax}
\newmm{\cagg}{C_{\ax\gamma}}
\newmm{\caggzero}{C_{\ax\gamma,0}^\text{NLO}}
\newmm{\gagg}{g_{\ax\gamma}}
\newmm{\gaee}{g_{\ax{}e}}
\newmm{\pag}{\prob_{\gamma \leftrightarrow a}}
\newmm{\afrac}{\eta_a}

\newcommand{\hyperlsw}{HyperLSW\xspace}
\newmm{\zopt}{\mtx{z}{opt}}
\newmm{\ngr}{\mtx{n}{g}}
\newmm{\nset}{\mtx{n}{s}}
\newmm{\nm}{N}
\newmm{\nref}{\mtx{n}{r}}
\newmm{\plaser}{P_\lambda}
\newmm{\gap}{\Delta}
\newmm{\siggap}{\sigma_{\gap}}
\newmm{\rgap}{\delta}

\newmm{\rholoc}{\rho_{\text{DM},\odot}}